\documentclass{aa}

\usepackage{graphicx}
\usepackage{epstopdf}
\usepackage{txfonts}
\usepackage[usenames,dvipsnames]{xcolor}
\usepackage{natbib}

\usepackage[colorlinks=true,linkcolor=blue,citecolor=magenta,pdfborder={0 0 0}]{hyperref}
\usepackage{amsmath}
\usepackage{multirow} 
\usepackage{booktabs} 

\newcommand{\galaxy}{Mrk~996}

\newcommand{\kms}{\mbox{km\,s$^{-1}$}}

\newcommand{\lya}{\relax \ifmmode {\mbox Ly}\alpha\else Ly$\alpha$\fi}
\newcommand{\ha}{\relax \ifmmode {\mbox H}\alpha\else H$\alpha$\fi}
\newcommand{\hg}{\relax \ifmmode {\mbox H}\gamma\else H$\gamma$\fi}
\newcommand{\hd}{\relax \ifmmode {\mbox H}\delta\else H$\delta$\fi}
\newcommand{\hb}{\relax \ifmmode {\mbox H}\beta\else H$\beta$\fi}
\newcommand{\sii}{\relax \ifmmode {\mbox S\,{\scshape ii}}\else S\,{\scshape ii}\fi}
\newcommand{\nii}{\relax \ifmmode {\mbox N\,{\scshape ii}}\else N\,{\scshape ii}\fi}
\newcommand{\neiii}{\relax \ifmmode {\mbox Ne\,{\scshape iii}}\else Ne\,{\scshape iii}\fi}
\newcommand{\oii}{\relax \ifmmode {\mbox O\,{\scshape ii}}\else O\,{\scshape ii}\fi}
\newcommand{\oi}{\relax \ifmmode {\mbox O\,{\scshape i}}\else O\,{\scshape i}\fi}
\newcommand{\oiii}{\relax \ifmmode {\mbox O\,{\scshape iii}}\else O\,{\scshape iii}\fi}
\newcommand{\hii}{\relax \ifmmode {\mbox H\,{\scshape ii}}\else H\,{\scshape ii}\fi}
\newcommand{\hi}{\relax \ifmmode {\mbox H\,{\scshape ii}}\else H\,{\scshape i}\fi}
\newcommand{\hmol}{\element[][][][2]{H}}
\newcommand{\lprimeco}{$\rm L^{\prime}_{CO}$}
\newcommand{\alphaco}{$\rm \alpha_{CO}$}
\graphicspath{{./}{figures/}}

\begin{document}
    
\title{Cold molecular gas distribution and kinematics in the low-metallicity dusty starburst of Mrk\,996 resolved with ALMA}
\titlerunning{CO content and kinematics of Mrk\,996}
\authorrunning{Slater et al.}
\author{
        R. Slater
            \inst{\ref{inst1},\ref{inst2}}
            \thanks{Corresponding author: \email{roy.slater@utalca.cl}}
        \and
        R. Amorín
            \inst{\ref{inst3}}
            \thanks{Corresponding author: \email{amorin@iaa.es}}
        \and
          J.A. Fernández-Ontiveros\inst{\ref{CEFCA}, \ref{CEFCA-UA}}
        \and
          F.J. Sáez-Ruiz\inst{\ref{CEFCA}} 
        \and
        M. S. Oey \inst{\ref{inst4}}
        \and
        B. L. James
        \inst{\ref{inst5}}
        \and
        M. Mingozzi
        \inst{\ref{inst5}}
        \and
        M. Llerena
        \inst{\ref{inst7}}
        \and
        M.G. del Valle-Espinosa
        \inst{\ref{inst6}}
        \and
        K. Harrington
        \inst{\ref{inst8},\ref{inst9},\ref{inst10}}
        \and
        N. Kumari
        \inst{\ref{inst5}}
        \and
        R. Sánchez-Janssen
        \inst{\ref{inst11}}
        \and
        J.M. Vílchez
        \inst{\ref{inst3}}
          }
\institute{Departamento de Tecnolog\'ias Industriales, Facultad de Ingeniería, Universidad de Talca, Los Niches km 1, Curic\'o, Chile \label{inst1}
    \and
    Departamento de F\'isica y Astronom\'ia, Universidad de La Serena, Avda. Juan Cisternas 1200, La Serena, Chile \label{inst2}
    \and
    Instituto de Astrofísica de Andalucía, CSIC, Apartado de correos 3004, E-18080 Granada, Spain  \label{inst3}
    \and
    Centro de Estudios de F\'isica del Cosmos de Arag\'on (CEFCA), Plaza San Juan 1, 44001 Teruel, Spain \label{CEFCA}
    \and
    Unidad Asociada CEFCA--IAA, CEFCA, Unidad Asociada al CSIC por el IAA y el IFCA, Plaza San Juan 1, 44001 Teruel, Spain \label{CEFCA-UA}
    \and
    Department of Astronomy, University of Michigan, Ann Arbor, MI 48109, USA. \label{inst4}
    \and
    ESA for AURA Space Telescope Science Institute 3700 San Martin Drive, Baltimore, MD 21218, USA) \label{inst5}
    \and
    Space Telescope Science Institute, 3700 San Martin Drive, Baltimore, MD 21218, USA \label{inst6}
    \and
    INAF - Osservatorio Astronomico di Roma, Via di Frascati 33, 00078, Monte Porzio Catone, Italy \label{inst7}
    \and   
    European Southern Observatory, Alonso de Córdova 3107, Vitacura, Casilla 19001, Santiago de Chile, Chile \label{inst8}
    \and
    Joint ALMA Observatory, Alonso de Córdova 3107, Vitacura, Casilla 19001, Santiago de Chile, Chile \label{inst9}
    \and
    National Astronomical Observatory of Japan, Los Abedules 3085 Oficina 701, Vitacura 763 0414, Santiago, Chile \label{inst10}
    \and
    Isaac Newton Group of Telescopes, Apartado 321, E-38700, Santa Cruz de la Palma, Tenerife, Spain \label{inst11}
    }

   \date{Received September XX, XXXX; accepted March XX, XXXX}


  \abstract
{Detecting cold molecular gas in metal-poor galaxies remains a major challenge. Their low carbon and oxygen abundances hinder carbon monoxide (CO) formation, while the low dust content reduces shielding against UV photodissociation in regions of intense star formation. As a result, CO, the main tracer of molecular gas, becomes faint or even undetectable in these environments.}
{We study the spatial distribution and kinematics of cold molecular gas in Mrk\,996, a nearby low-mass Wolf–Rayet galaxy hosting a dense, low-metallicity ($\sim$\,20\% solar) and nitrogen-enriched nuclear starburst with a complex ionized gas kinematics.}
 {Using ALMA observations of CO(1--0) and CO(2--1) at $\sim$113 pc arcsec$^{-1}$ and 1.3\,km\,s$^{-1}$ spectral resolution, we map the morphology and kinematics of the molecular gas and compare them with optical and UV data tracing the ionized gas and young stellar populations.}
{We detect compact CO clouds within $\sim$800\,pc of the starburst, spatially offset from the nuclear super star cluster (SSC) and the most highly ionized regions. The CO lines are narrow and supersonic, and exhibit a velocity gradient with a mild global blueshift, indicating dynamically perturbed gas without evidence for fast outflows, in contrast with the highly ionized phase. The global CO(2--1)/CO(1--0) ratio is low ($R_{21}\!\approx\!0.29$--0.34), consistent with 
subthermal excitation. 
The millimeter continuum peaks at the SSC, while CO emission is displaced toward obscured regions, suggesting that it traces dense shielded clumps. ALMA recovers about half of the single-dish flux, suggesting the presence of extended, low-surface-brightness molecular gas.  Using a metallicity-dependent CO-to-H$_2$ conversion factor, we infer a molecular gas mass of a few $10^{7}\,M_\odot$.}  
{The molecular gas is only weakly coupled to the stellar feedback that dominates the ionized phase. Our results support a multiphase scenario in which dense molecular clumps survive in shielded regions, while CO is photodissociated in their envelopes, leaving a significant  CO-dark H$_2$ component. }


   \maketitle
%

\section{Introduction}
Cold molecular hydrogen (\hmol) is the main reservoir of star-forming material in galaxies \citep[e.g.,][]{leroy2008,Bigiel2008,kennicutt2012}. However, because H$_2$ lacks a permanent dipole moment, it can only be observed directly in its warm phase ($T \gtrsim 100$\,K), leaving most of the cold molecular component ($T \sim 10$--20\,K) effectively invisible. For this reason, carbon monoxide (CO) is commonly used as an indirect tracer of molecular gas through the CO-to-\hmol\ conversion factor, $\alpha_{\rm CO}$ \citep[e.g.,][]{solbout2005,bolatto2013}.

In low-metallicity star-forming dwarf galaxies, however, CO becomes particularly difficult to detect and $\alpha_{\rm CO}$ depends strongly on metallicity \citep{schruba2012,hunt2015,amorin2016,Cicone2017,madden2020,Zhou2021,Henkel2022}. Below roughly the metallicity of the Small Magellanic Cloud ($\sim$20\% solar), the reduced carbon and oxygen abundances limit CO formation, while the low dust content weakens shielding against UV photodissociation \citep{wolfire2010,glover2011,narayanan2012,bolatto2013}. As a result, CO often traces only the densest and best-shielded fraction of the molecular medium, while a substantial fraction of H$_2$ remains CO-dark and must be inferred from alternative tracers such as dust or far-infrared cooling lines \citep[e.g.,][]{Rubio2015,gong2018,cormier2019,madden2020,Shi2020}. Understanding how CO-bright and CO-dark molecular gas are distributed in these systems is therefore essential for interpreting the cold gas budget and star formation under chemically primitive conditions.

The problem becomes even more interesting in compact low-metallicity starbursts hosting super star clusters (SSCs), where stellar feedback is both intense and spatially concentrated \citep[e.g.,][]{hunter24araa}. SSCs, with stellar masses above $10^5\,M_\odot$ packed within only a few parsecs, are expected to have a major impact on their surrounding interstellar medium (ISM). They form in dense molecular clouds, where short free-fall times and high local star-formation efficiencies may allow part of the natal material to remain bound and survive early feedback \citep[e.g.,][]{Kruijssen2012}. Yet the degree to which molecular gas is dispersed, accelerated, compressed, or photodissociated by young SSCs remains poorly constrained, especially at low metallicity.

Classical models predict that stellar winds and supernova ejecta thermalize inside SSCs and drive overpressured bubbles and superwinds \citep[e.g.,][]{Chevalier1985,mclow-mccray1988}. In the youngest and densest clusters, photoionization and radiation pressure can also contribute significantly to the feedback budget \citep[e.g.,][]{Krumholz2009}. A key open question is therefore how the molecular phase responds to this feedback. If molecular clouds are efficiently accelerated or disrupted, one may expect broad CO lines, strong velocity gradients, or outflow signatures. If, instead, the dense molecular gas remains only weakly coupled to the feedback, CO may survive in compact shielded clumps that are spatially offset from the ionizing sources and show only modest kinematic perturbations. Spatially resolved multiwavelength studies are required to distinguish between these possibilities.

The \textit{Atacama Large Millimeter Array} (ALMA) is uniquely suited to investigate these questions in nearby starbursts on the spatial scales relevant to SSC feedback. However, despite substantial observational effort over the last decade, only a limited number of low-mass starbursts with metallicities below $\sim$20\% solar have been mapped in CO with ALMA \citep[e.g.,][]{Cormier2017,cormier2019,Turner2015,Turner2017,Miura2018,Kepley2018,Consiglio2016,Schruba2017,Oey2017,Beck2018,madden2020,hunt2023}. These studies reveal compact and often clumpy CO-bright clouds embedded in environments where a large fraction of the molecular gas is likely CO-dark \citep[see][for a review]{hunter24araa}. They also suggest that the molecular, ionized, and dusty phases need not be spatially or kinematically coincident, particularly in galaxies dominated by strong feedback and patchy obscuration. Expanding this picture to a wider variety of nearby metal-poor starbursts is important for understanding both local dwarf galaxies and their connection to high-redshift star-forming systems \citep[e.g.,][]{madden2020,berg2022,mingozzi2022}.

In this context, Mrk\,996 is an especially interesting target. It is a nearby ($z=$\,0.0054; $D \simeq 23$\,Mpc) blue compact dwarf galaxy with a peculiar combination of a compact starburst core and an evolved dwarf elliptical-like host \citep{Papaderos96,Cairos2001,gildepaz2005,Amorin2009,thuan2008}. Its gas-phase metallicity is low ($12+\log({\rm O/H}) \sim 7.5$--8.3; roughly $\sim$0.2\,Z$_\odot$ on average), and nearly all current star formation is concentrated within a compact $\sim$3$^{\prime\prime}$ ($\sim$339\,pc) nuclear region hosting an SSC. This core reaches extreme electron densities of order $\sim10^{6}$ cm$^{-3}$ \citep{jaiswal2013,thuan2008}, far above those of typical \hii\ regions \citep{izotov1994} and comparable low-mass star-forming galaxies \citep[e.g.,][]{hunthira2009}. It also shows very broad optical emission lines (${\rm FWHM}\sim$400--900 km s$^{-1}$), strong high-excitation features, prominent Wolf--Rayet signatures, and a large nitrogen enhancement \citep{james2009,telles2014}, all pointing to an extreme feedback-dominated nuclear environment.

A similar combination of low metallicity, strong excitation, and enhanced N/O is now being identified in some compact galaxies observed with \textit{JWST}, which reinforces the value of Mrk\,996 as a nearby analogue of more extreme star-forming systems at high redshift \citep[e.g.,][]{Charbonnel2023,Senchyna2024,MarquesChaves2024,castellano2024,Arellano-Cordova2025}. In Mrk\,996, however, X-ray and optical diagnostics disfavor an AGN \citep{georgakakis2011,latimer2019}, indicating that the observed excitation and enrichment are more naturally explained by stellar feedback.

Despite its low metallicity and hard radiation field, Mrk\,996 has been detected in CO with single-dish observations \citep{hunt2015,hunt2017}. These measurements demonstrate the presence of cold molecular gas, but they are spatially unresolved and leave open key questions about its morphology, excitation, and kinematics, as well as its relation to the SSC, the ionized gas, and the dusty ISM. In particular, it remains unclear whether the molecular gas is strongly affected by the nuclear feedback, whether it shares the kinematics of any of the ionized gas components, or whether it survives mainly in shielded clumps displaced from the starburst.

In this paper we present new ALMA observations of the CO(1--0) and CO(2--1) transitions in Mrk\,996, combined with optical and UV data from \textit{HST} and Gemini/GMOS. Our main goal is to characterize the spatial distribution and kinematics of the cold molecular gas on sub-kiloparsec scales and to assess how it is coupled to the different phases of the ISM in this extreme low-metallicity starburst. Throughout the paper, we use these data to test whether the CO-emitting gas is decoupled from the nuclear feedback, or instead retains signatures of partial dynamical coupling while remaining confined to dense, shielded structures.

The paper is organized as follows. Section~\ref{sect-obs} describes the ALMA observations and ancillary datasets. Section~\ref{sect-res} presents the CO morphology, continuum emission, and molecular gas estimates. Section~\ref{sect-discuss} discusses the relation between the molecular and ionized gas phases, the nature of the CO reservoirs, and the implications for stellar feedback in Mrk\,996. Section~\ref{sect-concl} summarizes our main conclusions.

\begin{figure}[t!]
 \centering
 \includegraphics[bb=0 0 950 590,width=0.48\textwidth,clip]{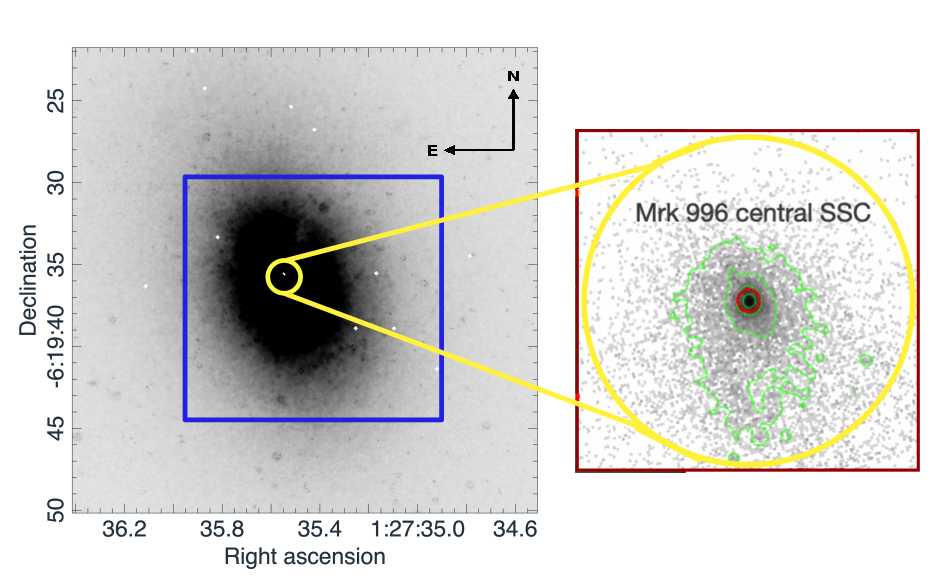}
   \caption{Left panel: combined HST (V+I-band) image of \galaxy\ \citep{thuan96}. The blue squared region represents the observed ALMA Field of View (FOV), clipped at 15\arcsec$\times$15\arcsec\ in the moment maps of Fig.~\ref{fig-moms}, with North to the top and East to the left. Right panel: the SSC from extremely compact NUV emission (yellow circle region) in the inner 1\arcsec radius, imaged by HST-COS acquisition (FWHM$\sim$0.1\arcsec\ ($\sim$11 pc), red circle). Green contours correspond four surface brightness levels, starting at 1$\sigma$ ($\sim$0.02 count s$^{-1}$), 3$\sigma$, and then increasing logarithmically by a factor of 6.}
   \label{fig1}
\end{figure}

\section{Observations}
\label{sect-obs}

\subsection{ALMA data}
\label{subsect-alma}
Observations of  \galaxy\ were carried out with ALMA during Cycle 8 (project ID: 2021.1.01307.S; PI: Amorín) to map the distribution and kinematics of the cold molecular gas on sub-kiloparsec scales. The adopted array configuration provides a spatial resolution of $\sim$0\farcs5 ($\sim$57\,pc), allowing us to probe the central $\sim$500--1000\,pc region hosting the nuclear starburst (Fig.~\ref{fig1}).

Two execution blocks were obtained on 2021 December 18 (44 antennas) and 2022 May 7 (42 antennas), targeting the CO(1--0) and CO(2--1) transitions in Bands 3 and 6, respectively. The CO(1--0) line (115.271\,GHz rest frequency) was observed with a total on-source time of 30.9\,min, a bandwidth of 1.875\,GHz, and a spectral resolution of $\sim$2.6\,km\,s$^{-1}$. The CO(2--1) transition (230.538\,GHz) was observed for 11.1\,min with the same bandwidth and a spectral resolution of $\sim$1.27\,km\,s$^{-1}$. The quasar J0006$-$0623 was used as bandpass and flux calibrator, while J0110$-$0741 and J0141$-$0928 served as phase calibrators for CO(1--0) and CO(2--1), respectively.

Data calibration was performed using the standard ALMA pipeline within CASA (versions 6.2.1.7 and 6.5.5.21; \citealt{mcmullin2007}). Imaging was carried out using the \texttt{tclean} task. We explored different weighting schemes (natural and uniform) and robust parameters to optimize the recovery of faint extended emission. The final datacubes adopt Briggs weighting with robust=0.5 and a Gaussian $uv$-taper, providing a good compromise between sensitivity and angular resolution. 
The resulting synthesized beams are $0\farcs87 \times 0\farcs80$ for CO(1--0) and $1\farcs21 \times 1\farcs04$ for CO(2--1), with rms noise levels of $\sim$1.6 and 2.3\,mJy\,beam$^{-1}$ per channel, respectively, measured in line-free regions.

Continuum subtraction was performed in the image domain using \texttt{imcontsub}, which provides stable results when the continuum is weak compared to the line emission. Continuum maps were generated using multifrequency synthesis (\texttt{mfs}) including all line-free channels.
Moment maps were constructed using \texttt{immoments} after primary-beam correction. To ensure robust detections, we applied thresholds of 3$\sigma$ for CO(1--0) and 4$\sigma$ for CO(2--1). Additional details on calibration, imaging, and masking strategies are provided in Appendix~A.

\begin{table}[t] 
    \centering
    \caption{Basic properties from literature for \galaxy.}\label{table-basics}
    \begin{tabular}{|lccc|}
    \hline
  \multicolumn{1}{|c|}{Data} &
  \multicolumn{1}{c|}{Value} &
  \multicolumn{1}{c|}{Notes} &
  \multicolumn{1}{c|}{References} \\
\hline
 R.A. & $\rm 01^{h}27^{m}35.51^{s}$  & \tablefootmark{(a)} & (1) \\
 Dec. & $\rm -06\degr19\arcmin36\farcs06$  & \tablefootmark{(a)} & (1) \\
 z    & 0.00541 & \tablefootmark{(b)} & (1)  \\
 D [Mpc] & 23.30$\pm$1.81 & \tablefootmark{(c)} & (2) \\
 V$_{hel}$ [km s$^{-1}$] & 1622$\pm$10 & \tablefootmark{(d)} & (3) \\
 E(B--V)$_{gas}$ & 0.50$\pm$0.01 & \tablefootmark{(e)} & (4) \\
n$_{e}$([\sii]) [cm$^{-3}$] & 408$\pm$40 & \tablefootmark{(f)} & (5) \\
 12+log(O/H)$_{\rm SDSS}$ & 7.53$\pm$0.08 & \tablefootmark{(f)} & (5) \\
  12+log(O/H)$_{\rm IFU}$ & 7.94 (8.38) $\pm$0.30 & \tablefootmark{(g)} & (6, 7) \\
 log(M$_{\star}$)$_{\rm SED}$ [$\rm M_{\odot}$] & 8.74$\pm$0.15 & \tablefootmark{(h)} & (4) \\
 log(M$_{\star}$)$_{\rm NIR}$ [$\rm M_{\odot}$] & 9.18 & \tablefootmark{(h)} & (8) \\
 SFR$_{\ha}$ [$\rm M_{\odot} yr^{-1}$] & 0.18 & \tablefootmark{(i)} & (8) \\[1mm] 
 log(M$_{\ion{H}{I}}$) [$\rm M_{\odot}$] & 8.00 & \tablefootmark{(j)} & (7) \\[1mm] 
 $\nu^{(1-0)}_{obs}$ [GHz] & 114.651 & \tablefootmark{(k)} & (9) \\
 $\nu^{(2-1)}_{obs}$ [GHz] & 229.298 & \tablefootmark{(k)} & (9) \\
 R.A.$_{obs.}$ & $\rm 01^{h}27^{m}35.52^{s}$ & \tablefootmark{(l)} & (9) \\
 Dec.$_{obs.}$ & $\rm -06\degr19\arcmin35\farcs44$ & \tablefootmark{(l)} & (9) \\
\hline
    \end{tabular}
    \tablefoot{\tablefoottext{a}{Reference coordinates (J2000) for ALMA observations}; \tablefoottext{b}{Reported spectroscopic redshift; \tablefoottext{c}{Luminosity distance assuming H$_{0}$ = 70.0 km s$^{-1}$ Mpc$^{-1}$, $\rm \Omega_{matter}$= 0.308 and $\rm \Omega_{\Lambda}$ = 0.692}}; \tablefoottext{d}{Central heliocentric velocity by \hi\ emission}; 
\tablefoottext{e}{Nebular reddening}, \tablefoottext{f}{[\sii] electron density and $T_e$-based oxygen abundance within the central 3$''$ ($\sim$ 339 pc)}
; \tablefoottext{g}
{Oxygen abundance in the \ha\ emitting region from spatially resolved spectra}; \tablefoottext{h}{Stellar mass estimated from SED fitting and near infrared luminosities}; \tablefoottext{i}{Star formation rate estimated from \ha\ and 24 $\mu$m luminosities}; \tablefoottext{j}{Total \hi\ mass, over entire \hi\ extent}; \tablefoottext{k}{Observed frequency of our ALMA observations corresponding to the adopted redshift.} ; \tablefoottext{l}{Central coordinates (J2000) of our ALMA observations, and coinciding with the CO continuum peaks.}  }
    \tablebib{(1) NASA/IPAC Extragalactic Database (NED, \url{http://ned.ipac.caltech.edu}); (2)\citealt{wright2006}; (3) \citealt{thuan96}; 
    (4) \citealt{berg2022}; (5) 
    \citealt{Arellano-Cordova2024}; (6) 
    \citealt{telles2014}; (7) \citealt{james2009}; (8) \citealt{hunt2015}; (9) This work.}
\end{table}

\subsection{Ancillary data}
\label{subsect-ancill}

\begin{figure}[t!]
 \centering
 \includegraphics[bb=55 0 605 260,width=0.45\textwidth,clip]{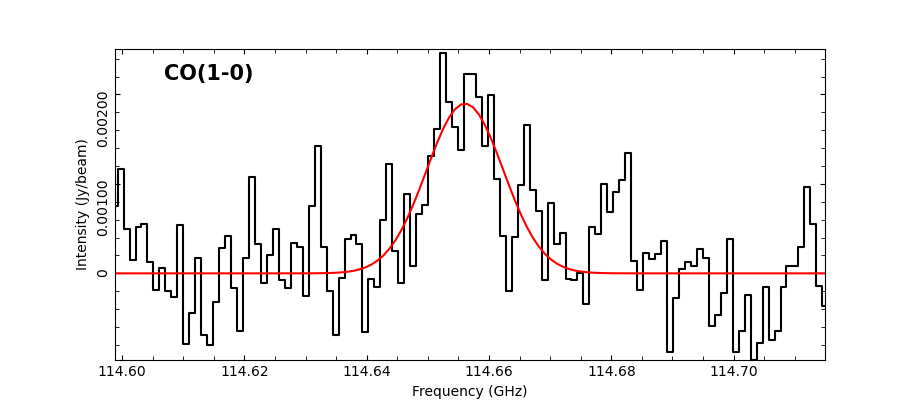}
 \includegraphics[bb=55 0 605 260,width=0.45\textwidth,clip]{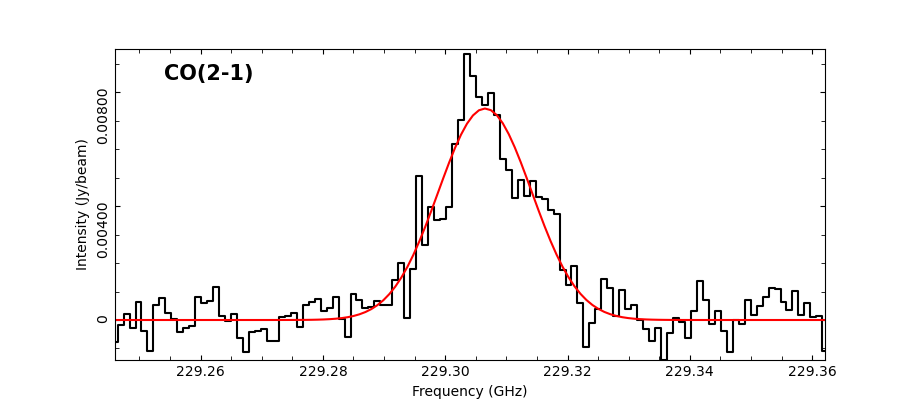}
   \caption{Averaged CO(1–0) and CO(2–1) spectra extracted over the central $15^{\prime\prime}$ region. Each spectrum was continuum-subtracted, primary-beam corrected, and derived from regions identified as genuine emission. Red lines show single-component Gaussian fits used to define the velocity ranges for the moment maps.   }
   \label{fig-spectCOs}
\end{figure}

To complement the ALMA observations, we used optical and ultraviolet data tracing the ionized gas and young stellar populations.
We retrieved near-ultraviolet (NUV) acquisition images from the \textit{Hubble Space Telescope} (HST)/COS as part of the CLASSY survey \citep{berg2022,james2022}, together with archival HST/WFPC2 optical images in the F569W and F791W filters \citep{thuan96}, approximately corresponding to the $V$ and $I$ bands. These data were used to trace the spatial distribution of the stellar continuum and the ionizing radiation field in the central region.

We also used integral-field spectroscopy from archival Gemini/GMOS \citep{telles2014}, covering the wavelength range 3667--7223\,\AA, to map the ionized gas morphology, excitation, and kinematics. Emission-line maps were constructed adopting a minimum signal-to-noise ratio of $\sim$10 for most lines, and $\gtrsim$8 for the [O\,II] $\lambda\lambda3726,3729$ doublet.

Finally, we also used archival VLT/VIMOS integral-field data presented by \citet{james2009}, from which we derived the H$\alpha$ contours shown in Sect.~\ref{sect-discuss}. In addition, we used archival VLT/MUSE narrow-field mode adaptive-optics data (ID: 114.27GX, PI: R. Amorín), presented in a separate work (Mingozzi et al., in prep.), to trace the morphology of the ionized gas at higher angular resolution. Both datasets were reoriented and astrometrically registered to the Gaia reference frame, yielding an astrometric solution consistent with that of the ALMA data. These datasets provide a direct comparison between the cold molecular gas traced by CO and the ionized and stellar components of the interstellar medium, which is central to the analysis presented in Sect.~\ref{sect-discuss}.

\section{Results}
\label{sect-res}

The CO emission lines are detected at signal-to-noise ratios above 2 per channel across the central $\sim$15$^{\prime\prime}$ ($\sim$1.7\,kpc) of the ALMA field of view (FOV). The emission is spatially structured and clumpy, with distinct components that differ in morphology, excitation, and kinematic properties. In both transitions, the CO emission is characterized by relatively narrow line profiles but non-negligible velocity offsets and gradients, indicating that the molecular gas exhibits coherent large-scale motions and is mildly perturbed on sub-kiloparsec scales.

In CO(1–0), the signal appears diffuse and extended, with a prominent region northwest of the galaxy center and additional faint substructures, including a tiny emission feature nearly coincident with the SSC, slightly offset to the north.
In CO(2–1), two clearly separated components are observed: a bright cloud towards the northwest that coincides with the CO(1–0) cloud and connects with a southern elongated cloud near the nucleus, and a smaller cloud towards the farthest southwest ($\sim$ 5\arcsec from the SSC), both with peak SNRs reaching up to $\sim$10.
Figure~\ref{fig-spectCOs} displays the integrated spectra for both transitions. Gaussian fits mark the spectral range used to construct the moment maps (114.64-114.67\,GHz for CO(1–0) and 229.27-229.33\,GHz for CO(2–1)).
\begin{figure*}[!t]
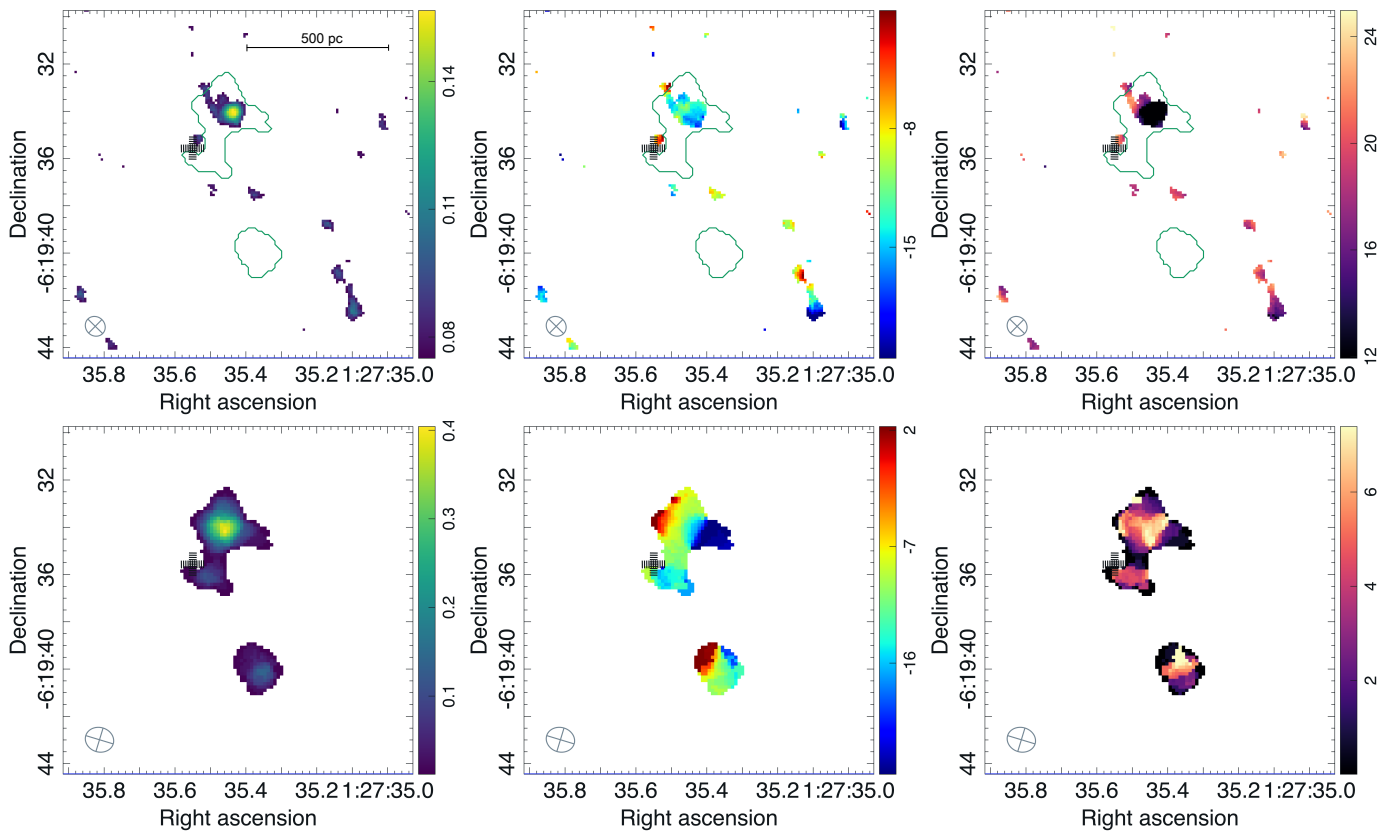

 \centering
 \includegraphics[width=0.99\textwidth]{figures/Mrk996_co10_moms_v2-edited.pdf}
 \includegraphics[width=0.99\textwidth]{figures/Mrk996_co21_moms_v2.pdf}
 \caption{Moment maps of CO(1–0) (top) and CO(2–1) (bottom) emission in \galaxy. All maps are shown with north up and east to the left, as in Fig.~\ref{fig1}. From left to right: integrated intensity scaled in Jy beam$^{-1}$ km s$^{-1}$, and both velocity and velocity dispersion in km s$^{-1}$. Contours correspond to the $4\sigma$ CO(2–1) flux level for reference. A dashed cross marks the position of the UV peak from the SSC. All maps are shown in relative coordinates centered on the ALMA pointing.}
   \label{fig-moms}
\end{figure*}

\subsection{Moment maps}
\label{subsect-mom}

Figure~\ref{fig-moms} shows the CO(1–0) and CO(2–1) integrated intensity (moment~0), velocity (moment~1), and observed velocity dispersion (moment~2) maps.
All maps are masked below $3\sigma$ (CO(1–0)) or $4\sigma$ (CO(2–1)) to ensure robust detections.
Unless stated otherwise, all velocities are measured relative to the systemic velocity listed in Table~\ref{table-basics}. For a direct comparison of the CO emission, the SSC, and the optical line emission, providing a clearer sense of spatial scale, see later Figs.~\ref{fig-hst} and \ref{fig-ratios1} and the discussion in Sect.~\ref{sect-discuss}.

\subsubsection{CO(1–0)}
\label{subsect-co10}
The CO(1–0) emission is concentrated in a compact region extending $\sim$3\farcs2 ($\sim$360\,pc) from the SSC, which we refer to as the \emph{Main Cloud} (MC$_{1-0}$).
A faint one-arm feature extends northeastward.
The maximum integrated flux reaches $\sim$0.16\,Jy\,beam$^{-1}$\,km\,s$^{-1}$, while secondary components remain below 70\% of that value.
The velocity field shows mean velocities of $\sim$$-12.5$\,km\,s$^{-1}$ with a total range of 0 to $-25$\,km\,s$^{-1}$. No rotational signature is detected, and the velocity pattern appears irregular and asymmetric.

The velocity dispersions range from $\sim$9 to 23\,km\,s$^{-1}$, increasing from the central regions toward the northeastern extension. Although these values are relatively low compared to the highly ionized gas, they are clearly supersonic for cold molecular gas, implying the presence of significant turbulence. The observed dispersions are broadly consistent with gravitationally bound or marginally bound structures, but they also suggest that the molecular gas is subject to external perturbations, likely related to the feedback processes in the central starburst.

\subsubsection{CO(2–1)}
\label{subsect-co21}
The CO(2–1) emission is brighter and detected over a larger area than CO(1–0), spanning $\sim$10$^{\prime\prime}$ ($\sim$1.13\,kpc), although this difference may be partly influenced by the higher signal-to-noise ratio of the CO(2–1) data.
Two distinct components are resolved:
(1) a large, irregular \emph{Main Cloud} (MC$_{2-1}$) surrounding the optical nucleus and with a peak emission roughly coincident with the peak of the MC$_{1-0}$, and 
(2) a smaller, roughly spherical \emph{Secondary Cloud} (SC$_{2-1}$) located $\sim$4$^{\prime\prime}$ ($\sim$452 pc) southwest of the SSC.
Peak fluxes reach $\sim$0.41\,Jy\,beam$^{-1}$\,km\,s$^{-1}$ for the MC$_{2-1}$ and $\sim$0.14\,Jy\,beam$^{-1}$\,km\,s$^{-1}$ for the SC$_{2-1}$.
In the region where both CO transitions overlap, the morphology and radial velocities are in good agreement (see middle panels of Fig.~\ref{fig-moms}).

The CO(2--1) velocity field spans from $\sim -27$ to $+9$\,km\,s$^{-1}$, with mean velocities around $-12.8$\,km\,s$^{-1}$. A coherent velocity gradient is observed across the MC$_{2-1}$, although no clear large-scale rotational pattern is evident. The velocity dispersions are low, ranging from $\sim$3 to 12\,km\,s$^{-1}$, but remain supersonic, indicating that the gas is turbulent despite the absence of strong line broadening. These properties indicate that the molecular gas displays coherent bulk motions superimposed on a turbulent velocity field.

\subsection{Continuum maps}
\label{subsect-cont}

\begin{figure}[ht!]
 \centering
 \includegraphics[bb=0 0 450 400,width=0.9\columnwidth,clip]{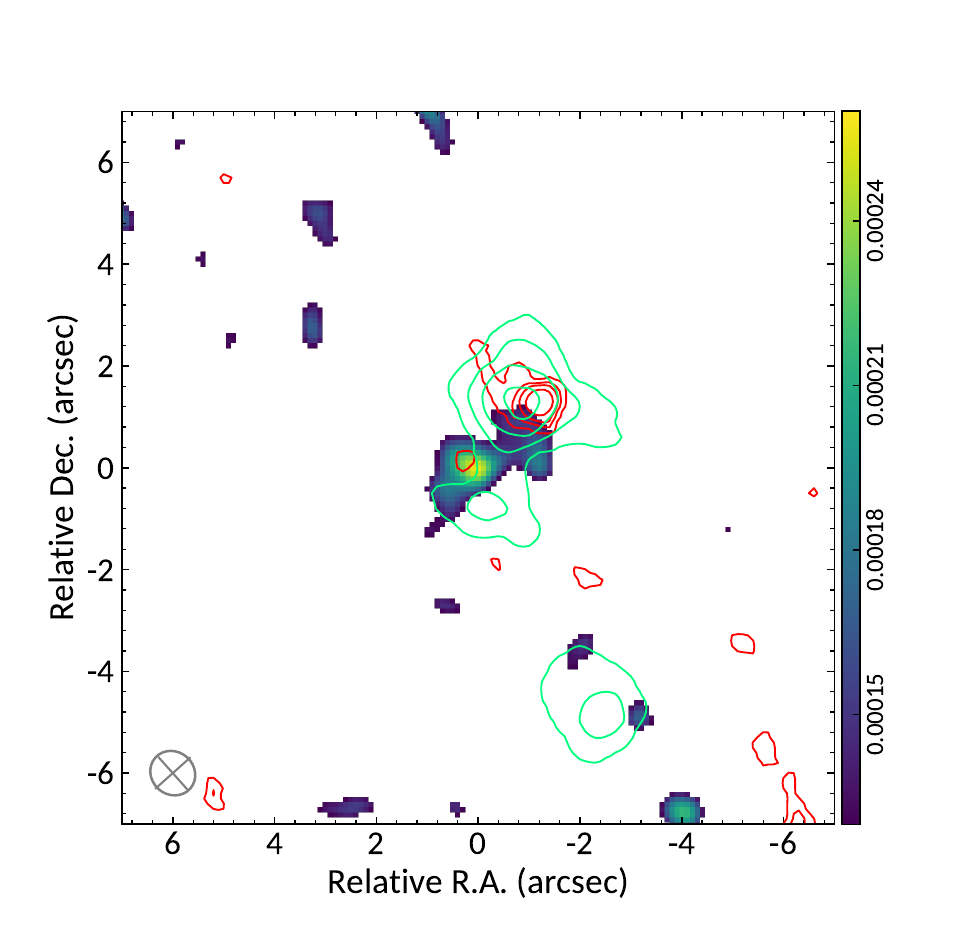}
 \includegraphics[bb=0 0 450 400,width=0.9\columnwidth,clip]{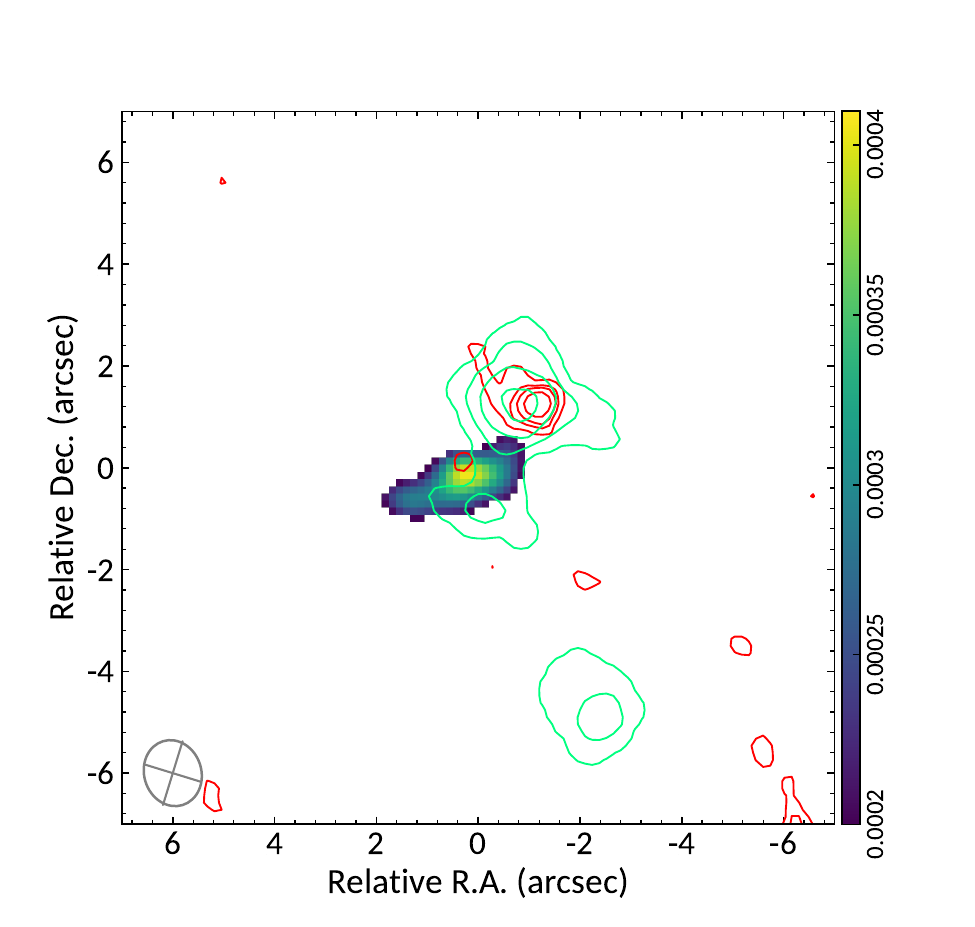}
   \caption{Continuum maps at both 115 GHz (Band-3) and 230 GHz (Band-6) scaled by a color bar in Jy beam$^{-1}$ (top and bottom panel, respectively). Red and green overlays, represent the contours of the 0th-moment above the minimum values of CO(1–0) (7.5$\times$10$^{-2}$ Jy beam$^{-1}$ \kms) and CO(2–1) (1.18$\times$10$^{-2}$ Jy beam$^{-1}$ \kms), respectively. The coordinates are relative to the center of our ALMA observations (Table~\ref{table-basics}), which matches the innermost region of the continuum emissions. In the 115 GHz map, pixels with values lower than 0.13 mJy beam$^{-1}$ were not considered ($\sim$2.5$\sigma$), whereas for the 230 GHz map, pixels lower than 0.2 mJy beam$^{-1}$ ($\sim$2$\sigma$).}
   \label{fig-conts}
\end{figure}

Figure~\ref{fig-conts} shows the 115 and 230\,GHz continuum maps. At these frequencies, the continuum emission is likely dominated by a combination of thermal dust emission and free--free emission, with a possible minor synchrotron contribution. 
The 230 GHz emission is expected to be mostly dominated by dust, while at 115 GHz, a non-negligible fraction may instead arise from free–free emission. 
At 115\,GHz, faint extended emission is detected out to $\sim$15\arcsec, with a compact component dominating the central 4\arcsec. 
At 230\,GHz, the continuum is detected entirely confined to the central 3\arcsec\ region, co-spatial with the SSC and the 115\,GHz peak. The emission is dominated by central components with flux densities of $\sim$0.60,mJy at 115,GHz and $\sim$0.47,mJy at 230,GHz.
The peak of the millimeter continuum is co-spatial with the small CO(1–0) clump in the immediate vicinity of the SSC. 
In that region, the 230\,GHz continuum is brighter than at 115\,GHz ($\approx$0.40 vs.\ $\sim$0.26\,mJy\,beam$^{-1}$), consistent with a larger contribution from warm dust and/or higher optical depth near the starburst. 

\begin{figure*}[t!]
  \centering
  \includegraphics[width=0.95\textwidth]{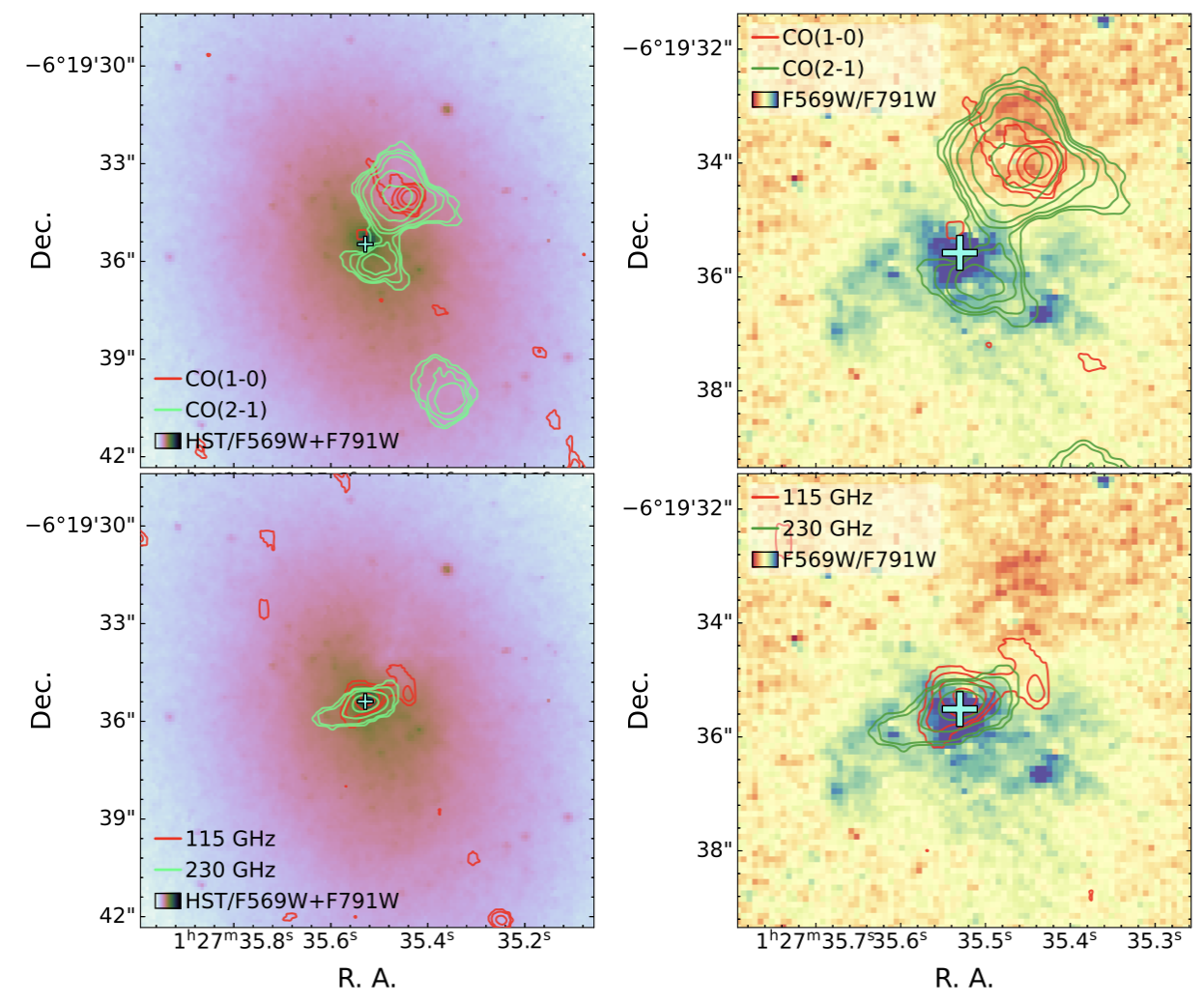}
  \caption{Optical view of \galaxy: (\textit{Left}) HST V+I map and (\textit{Right}) V/I color map. CO contours are overlaid in the top panels, while dust-dominated continuum contours at 115 GHz (band-3, red) and 230 GHz (band-6, green) are shown in the bottom panels. Contour levels are similar to those in Fig.~\ref{fig-conts}. Coloured scale bars were used for each image, with darker (bluer) colors indicating brighter and bluest regions, respectively.
  The bluest and brightest optical regions concentrate stellar and nebular emission in the nuclear SSC and its surroundings, co-spatial with the mm continuum emission but offset with CO-bright structures, consistent with the slight offset between warm dust and the bulk of the cold molecular gas.}
  \label{fig-hst}
\end{figure*}

For context, a combined HST image by summing the flux from V (F569W) and I (F791W) bands (V$+$I) and the V--I color map in Fig.~\ref{fig-hst} show that the bluest optical regions are concentrated around the nucleus but do not perfectly coincide with all CO-bright structures. The left panel of Fig.~\ref{fig-hst} highlights the optical morphology relative to the CO contours, while the right panel emphasizes the spatial overlap of the central burst and both 115 and 230\,GHz continuum contours. Together with the continuum maps, this indicates that the brightest dust mm emission is tightly associated with the SSC, whereas the bulk of the cold molecular gas traced by CO is slightly offset from both the UV/optical peaks and the dustiest optical patches. This spatial offset between the dust continuum peak and the CO-bright regions already suggests a complex internal structure of the ISM, where different phases respond differently to the central starburst. Further comparisons with ionized gas components from previous studies are presented in Sect.~\ref{sect-discuss}.

\subsection{Molecular gas estimations}
\label{subsect-molmass}

We estimated molecular gas masses from the integrated CO fluxes using the standard luminosity relation from \citet{solomon97}, 
\begin{equation}
L'_{\rm CO} = 3.25\times10^{7}\,
\left(\frac{S_{\rm line}\Delta v}{\rm Jy\,km\,s^{-1}}\right)
\left(\frac{D_L}{\rm Mpc}\right)^2
\left(\frac{\nu_{\rm obs}}{\rm GHz}\right)^{-2}
(1+z)^{-3},
\end{equation}
in K\,km\,s$^{-1}$\,pc$^2$, 
where $S_{\rm line}\Delta v$ is the integrated flux in Jy\,km\,s$^{-1}$, 
$D_L$ the luminosity distance (Mpc), and $\nu_{\rm obs}$ the observed frequency (GHz).
The molecular mass is then obtained as, 
\begin{equation}
M_{\rm H_2}=\alpha_{\rm CO}\,L'_{\rm CO},
\end{equation}
where $\alpha_{\rm CO}$ (in $M_\odot\,({\rm K\,km\,s^{-1}\,pc^2})^{-1}$) includes helium (i.e. $\alpha_{\rm CO,MW}=4.35$).
We adopted the metallicity-dependent conversion factor from \citet{amorin2016},
$\log(\alpha_{\rm CO}) = 0.7 - 1.5\,[(12+\log{\rm (O/H)})-8.7]$,
which yields $\alpha_{\rm CO}\!\approx\!69\,M_\odot\,({\rm K\,km\,s^{-1}\,pc^2})^{-1}$ for $12+\log({\rm O/H})=7.94$.

Table~\ref{table-prop} summarizes the resulting luminosities and gas masses.
The total molecular gas mass inferred from CO(1–0) is $M_{\rm H_2}\!\approx\!4.5\times10^{7}\,M_\odot$, about three times higher than that derived from CO(2–1), which may indicate that CO(1–0) traces a more extended and diffuse component.
We note that the dominant source of uncertainty in these estimates arises from the choice of $\alpha_{\rm CO}$, which may vary by $\sim$0.3--0.4 dex at this metallicity \citep{madden2020}. Systematic differences in molecular mass estimations due to the adoption of different $\alpha_{\rm CO}$ calibrations based on alternative methods and using slightly different metallicity values from literature are presented in Appendix~\ref{app:alphaCO}. 

\subsection{CO(2–1)/CO(1–0) ratio}
\label{subsect-r21}

To assess the excitation conditions, we computed the line luminosity ratio
$R_{21} = L'_{\rm CO(2-1)}/L'_{\rm CO(1-0)}$ \citep[e.g.,][]{saintonge2017,montoya2023}. This ratio is sensitive to the density and temperature of the molecular gas \citep[e.g.,][]{penaloza2018,denbrok2025}, and is often assumed to take a characteristic value in nearby galaxies \citep[e.g.,][]{koda2012,saintonge2017,keenan2024}. 

To ensure consistent sampling, the CO(2–1) moment-0 map was re-binned to match the pixel scale of the CO(1–0) map before measuring integrated fluxes only within the overlapping region between these transitions (more details, in Sect.~\ref{app:subsec-sub_mask}).
We obtained $L'_{\rm CO(1-0)}=(2.8\pm0.4)\times10^{5}$ and $L'_{\rm CO(2-1)}=(0.9\pm0.6)\times10^{5}$\,K\,km\,s$^{-1}$\,pc$^2$, resulting in $R_{21}=0.34\pm0.22$.
This value agrees with the ratio measured over the entire ALMA FOV (Table~\ref{table-prop}) and is significantly lower than the typical $R_{21}\!\sim\!0.6$1.0 observed in starburst dwarfs and spiral centers \citep[e.g.,][]{leroy2009,denbrok2021}.

The low $R_{21}$ value suggests subthermal excitation conditions, typically associated with low gas densities and/or beam dilution effects \citep[e.g.,][]{Crosthwaite2007}. In this regime, the $J=2$ level is not efficiently populated, leading to suppressed CO(2--1) emission \citep[e.g.,][]{eckart90,Israel2005}. However, the interpretation of $R_{21}$ is degenerate, as it depends on a combination of gas density, temperature, and filling factor \citep[e.g.,][]{penaloza2018,denbrok2025}. Therefore, the low ratio likely reflects a mixture of relatively low-density molecular gas and unresolved substructure within the beam.

\begin{table*}
	\centering
	\caption{Physical properties for molecular mass estimations in \galaxy.\tablefootmark{(a)}}
	\label{table-prop}
	\begin{tabular}{|c|c|c|c|c|c|}
		\hline
		\multicolumn{2}{|c|}{CO Region} &
		\multicolumn{1}{c|}{$S\Delta\nu$} &
        \multicolumn{1}{c|}{$\nu_{obs}$} &
		\multicolumn{1}{c|}{\lprimeco ($\times 10^{5}$)} &
		\multicolumn{1}{c|}{$\rm M_{gas}$ ($\times 10^{7}$)} \\
		\multicolumn{2}{|c|}{ } & [mJy km s$^{-1}$] & [GHz] & [$\rm K~km~s^{-1}~pc^{2}$] & [M$_{\odot}$] \\
		\multicolumn{2}{|c|}{(1)} & (2) & (3) & (4) & (5) \\
		\hline
		\multirow{2}{*}{CO(1–0)} & Total & $\rm 491.6 \pm 0.2$ & \multirow{2}{*}{114.656 $\pm$ 0.008} & $\rm 6.49 \pm 1.01$ & $\rm 4.49 \pm 2.46$ \\
		& Main Cloud & $\rm 226.9 \pm 0.3$ & & $\rm 3.00 \pm 0.47$ & $\rm 2.07 \pm 1.14$ \\
		\hline
		\multirow{3}{*}{CO(2–1)} & Total & $\rm 668.0 \pm 1.0$ & \multirow{3}{*}{229.303 $\pm$ 0.02} & $\rm 2.21 \pm 0.34$ & $\rm 1.53 \pm 0.84$ \\
		& Main Cloud & $\rm 563.0 \pm 1.0$ & & $\rm 1.86 \pm 0.29$ & $\rm 1.29 \pm 0.70$ \\
		& Small Cloud & $\rm 104.4 \pm 0.5$ & & $\rm 0.34 \pm 0.05$ & $\rm 0.24 \pm 0.13$ \\
		\hline
	\end{tabular}
	\tablefoot{\tablefoottext{a}{\alphaco\ $\approx$ 69.2 $\rm M_{\odot}~(K~km~s^{-1}~pc^{2})^{-1}$, by following \citet{amorin2016}. For different \alphaco\ prescriptions, see Appendix~\ref{app:alphaCO}.
	Columns: (1) Custom division for molecular estimations. "Total" means considering the entire field of view, as illustrated in Fig.~\ref{fig1} ;(2) integrated flux density; (3) observed frequency according to the peak line emission; (4) CO luminosity; (5) Molecular gas derived from CO luminosity and \alphaco. All errors are estimated by the error propagation method, except for flux density (standard deviation).}}
\end{table*}

\subsection{CO comparison with single-dish data}
\label{subsect-sdish}
To quantify potential flux losses due to interferometric filtering, we compared our measurements with single-dish observations from \citet{hunt2015,hunt2017}.
Those authors reported CO(1–0) and CO(2–1) intensities of 0.243 and 0.291\,K\,km\,s$^{-1}$ from IRAM 30\,m, corresponding to 1.2 and 1.4\,Jy\,km\,s$^{-1}$ after beam corrections (For details, see Sect.~\ref{app:subsec-fluxMRS}). Although CO(2–1) observations from APEX were also reported by these authors, they do not show a clearly defined line profile and may be affected by noise and instrument uncertainties, as also noted in \citet{hunt2017}. When compared to IRAM measurements, our ALMA data recover $\sim$42\% of the CO(1–0) flux and 50\% of the CO(2–1) flux. 
This is consistent with ALMA resolving out emission on scales larger than the maximum recoverable scale ($\sim$7$^{\prime\prime}$), 
indicating that part of the molecular gas is distributed in extended, low-surface-brightness structures that are filtered out by the interferometer. 
Accounting for the estimated flux recovery factors ($f_{1-0}\!\approx\!0.42$, $f_{2-1}\!\approx\!0.5$), the corrected CO line ratio 
becomes $R_{21}^{\rm corr}\approx 0.29$, which is substantially lower than 
single-dish values.
We summarize the measurements obtained with these datasets in Table~\ref{table-co_compare}. 
A brief discussion of typical R$_{21}$ values in the literature is provided in Sect.~\ref{app:subsec-fluxMRS}.

\begin{table*}[ht!]
\renewcommand{\arraystretch}{1.4}
\centering
\caption{Comparisons of total flux densities. Beam size or maximum recoverable scale (MRS) for each case, along with R$_{21}$ coefficients.}\label{table-co_compare}
\begin{tabular}{lcccc}
\toprule
\textbf{Instrument} & \textbf{Line} & \textbf{Total Flux Density [Jy km/s]} & \textbf{Resolution (\arcsec)} & \textbf{Notes} \\
\midrule
\multirow{2}{*}{IRAM 30m} & CO(1–0)  & \(1.2 \pm 0.2\)   & 21.5 (beam) & \multirow{2}{*}{Single-dish}\tablefootmark{(a)} \\
                          & CO(2–1)  & \(1.4 \pm 0.2\)   & 10.7 (beam) &                                 \\
\midrule
\multirow{2}{*}{ALMA}     & CO(1–0)  & 0.5               & 7.1 (MRS)   & \multirow{2}{*}{Interferometry} \\
                          & CO(2–1)  & 0.7               & 7.7 (MRS)  &                                 \\
\midrule
\midrule
\textbf{Ratio \( R_{21} \)} & IRAM     & 0.6               & \multicolumn{2}{l}{beam corrected to 22\arcsec} \\
                            & ALMA     & 0.34              & \multicolumn{2}{l}{see Sect.~\ref{subsect-r21}} \\
\midrule
\textbf{Ratio \( R_{21}^{corr} \)} & ALMA     & 0.29               & \multicolumn{2}{l}{corrected by flux recovery factors} \\
\bottomrule
\end{tabular}
\tablefoot{\tablefoottext{a}{\citealt{hunt2017}.}}
\end{table*}

%

\subsection{CO kinematics}
\label{subsect-dis_kin}
\label{sec:kinematics}

\begin{figure*}
 \centering
 \includegraphics[bb=10 0 2600 1450,width=0.9\textwidth,clip]{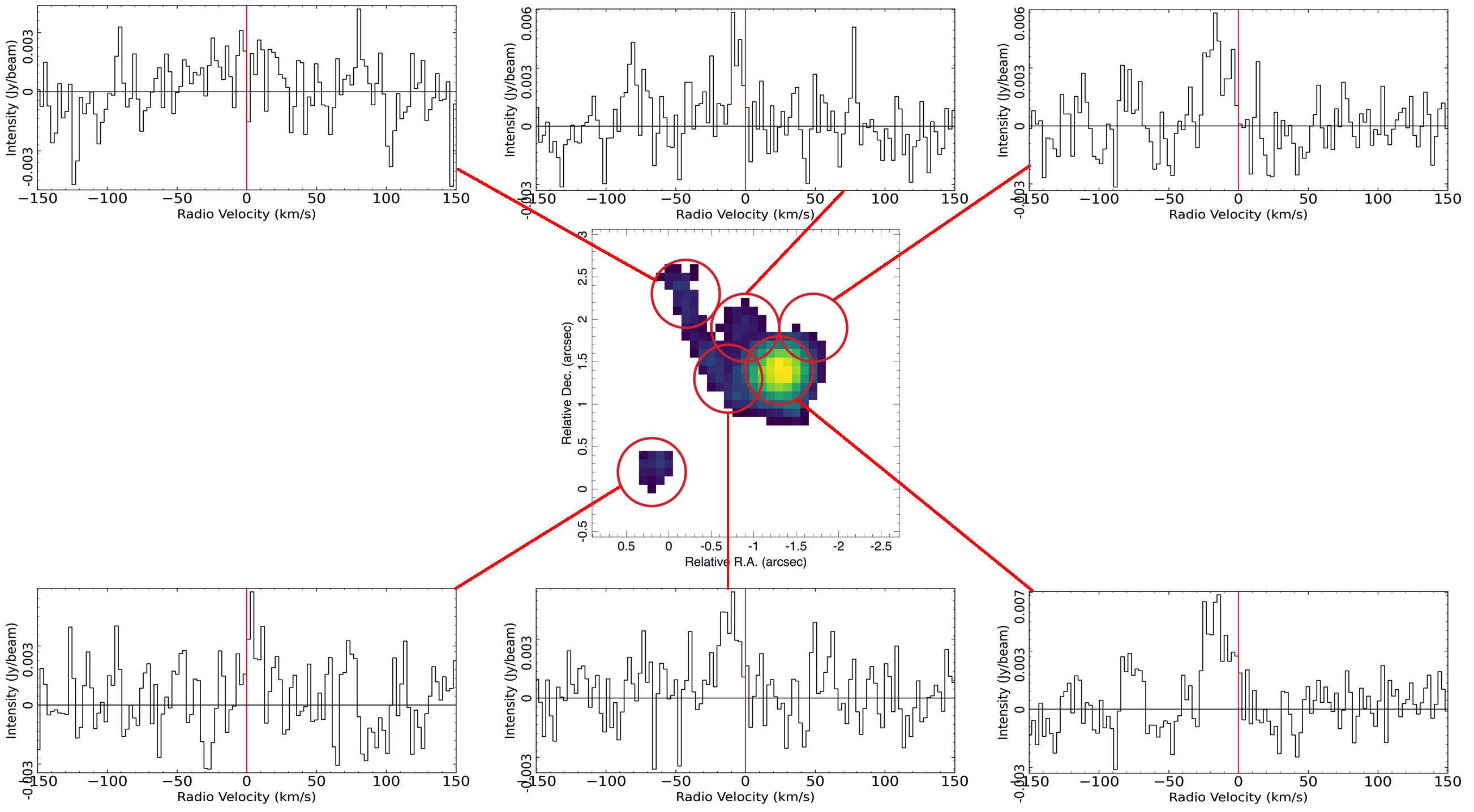}
   \caption{Representative CO(1–0) spectra extracted from 0\farcs4 apertures (roughly corresponding to the spatial resolution of our observations ) at the positions marked in the central maps. The vertical red line indicates the systemic velocity. 
   All profiles show relatively narrow linewidths and modest velocity offsets, consistent with mildly perturbed molecular gas. 
   The coordinates are relative to the center of our ALMA observations (Table~\ref{table-basics}).   }
   \label{fig-dis_co10}
\end{figure*}

\begin{figure*}[ht!]
 \centering
 \includegraphics[bb=10 0 2600 1450,width=0.9\textwidth,clip]{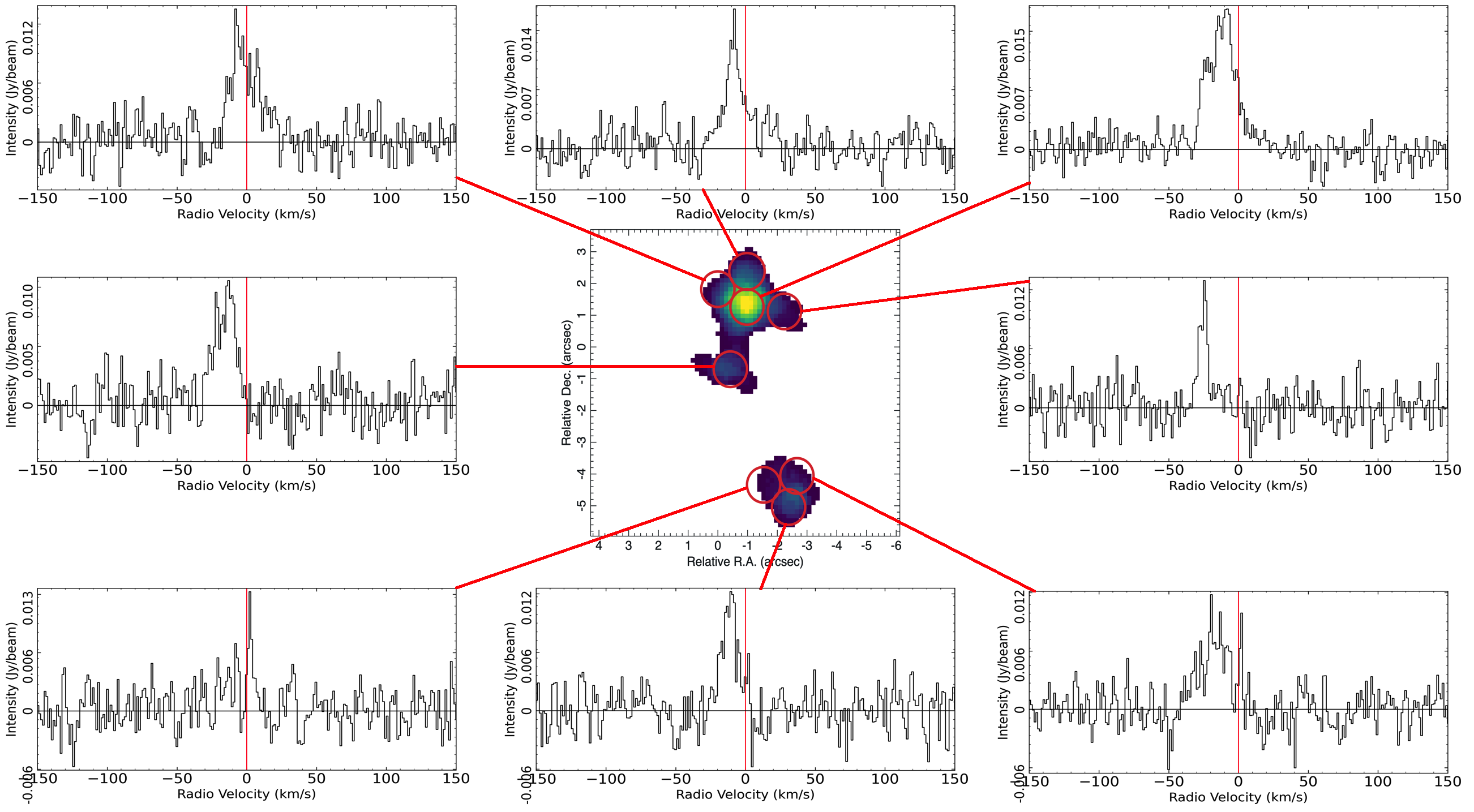}
   \caption{Same as in Fig.\ref{fig-dis_co10}, but considering the averaged spectra of the CO(2–1) emission line. As pointed out in Sect.~\ref{subsect-alma}, the spectral resolution for CO(2–1) transition is 1.27 \kms.
   }
   \label{fig-dis_co21}
\end{figure*}

Figures~\ref{fig-dis_co10} and ~\ref{fig-dis_co21} show representative CO spectra extracted from circular apertures (radius $0\farcs4$) across the main molecular structures. Both transitions exhibit relatively narrow line profiles (FWZI $\lesssim$ 80\,km\,s$^{-1}$) and small velocity offsets ($\lesssim$ 30\,km\,s$^{-1}$), with no clear evidence for broad wings. Notably, spectra extracted from the brightest regions of the two CO transitions, both co-spatial within the beam (Fig.~\ref{fig-conts}) and located $\sim$800 pc northwest of the SSC, show asymmetric profiles with a low-level red tail extending $\sim$25$\kms$ from the nebular systemic velocity. 
This red tail persists in the adjacent eastern region of the MC$_{2-1}$, suggesting a possibly slow molecular flow or shell expansion in a region where the ionized gas (H$\alpha$) no longer exhibits the broad components (FWHM $\sim$400-900\,km\,s$^{-1}$) seen in the nuclear region \citep{thuan96,james2009,telles2014}. 

Despite their narrow appearance, the measured velocity dispersions (3-12\,km\,s$^{-1}$ for CO(2--1) and 9-23\,km\,s$^{-1}$ for CO(1--0)) are supersonic for cold molecular gas ($T\sim10$-20 K), indicating turbulence driven by gravitational motions and/or external perturbations from the starburst. The CO velocity field shows a coherent gradient from W to E over a few hundred pc scales and a systematic blueshift peak emission.

Assuming virial equilibrium, the expected velocity dispersions are $\sigma_{\rm vir} \sim \sqrt{GM/R}$, yielding $\sim$30\,km\,s$^{-1}$ for CO(1--0) and $\sim$10\,km\,s$^{-1}$ for CO(2--1), for the characteristic sizes of the main molecular structures. These values are broadly consistent with the observed dispersions, supporting the interpretation that the gas resides in gravitationally bound or marginally bound clumps.

\section{Discussion}
\label{sect-discuss}

\subsection{Multiphase view of Mrk\,996}\label{subsect-comp_opt} 

\begin{figure}[t!]
 \centering
 \includegraphics[width=0.49\textwidth]{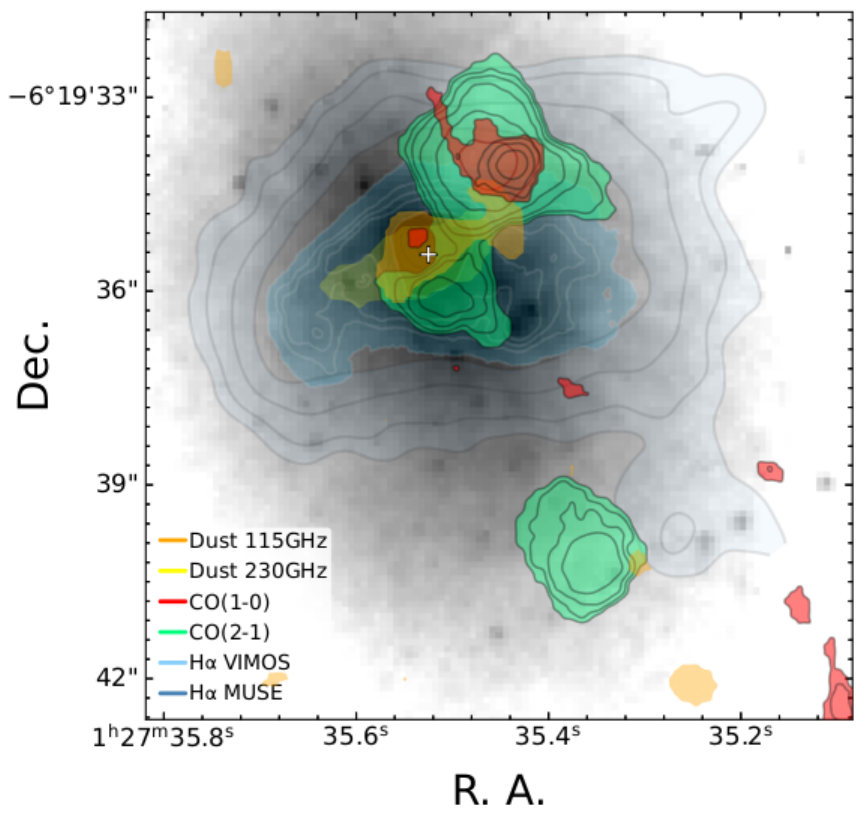}
 \caption{Multi-phase view of the central region of \galaxy. The background image shows the HST/WFPC2 $V+I$ (F569W+F791W) emission. Overlaid contours trace the molecular gas and cold dust from ALMA, together with the ionized gas from VLT/VIMOS H$\alpha$ \citep{james2009} and archival VLT/MUSE narrow-field mode adaptive-optics observations (Mingozzi et al., in prep.). The VIMOS and MUSE data were reoriented and astrometrically registered to the Gaia reference frame, yielding an astrometric solution consistent with that of the ALMA data. The figure highlights the spatial offset between the CO-bright molecular gas and the compact, highly ionized nuclear region associated with the SSC, while also showing the connection between the molecular gas and the dustier, more shielded regions.}
 \label{fig-multiphase}
\end{figure}

Figure~\ref{fig-multiphase} provides a direct multiwavelength view of the central region of \galaxy, combining the stellar continuum with tracers of the molecular gas, cold dust, and ionized gas. A clear spatial differentiation between ISM phases is observed. The CO-bright molecular component is not co-spatial with the compact ionized core powered by the SSC. Instead, the molecular gas is displaced toward more obscured structures to the northwest, as also seen in the optical continuum (Fig.~\ref{fig-hst}). 

\begin{figure*}[ht!]
 \centering
 \includegraphics[width=\textwidth]{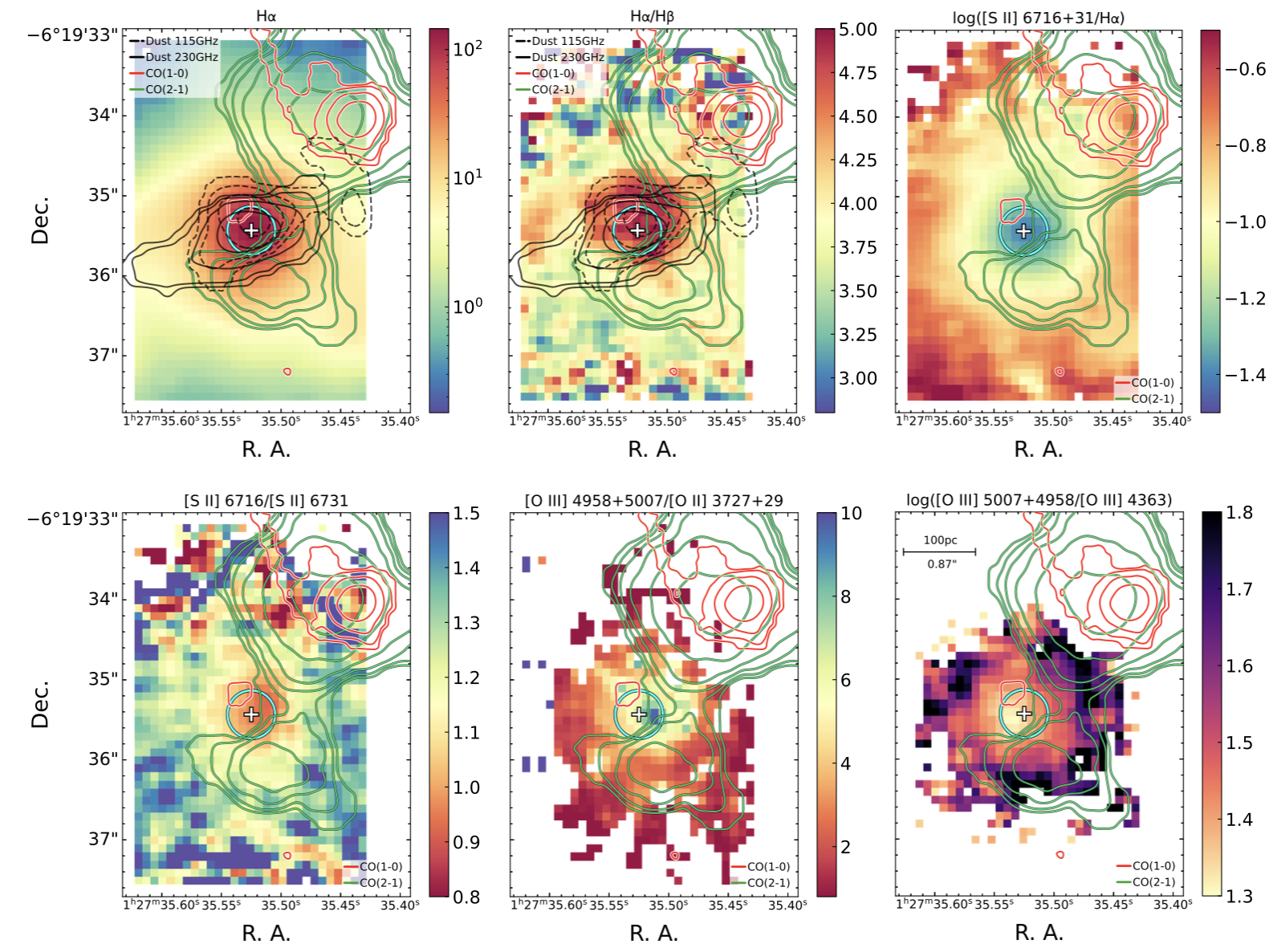}
 \caption{Optical emission-line ratio maps from the Gemini/GMOS IFU data of \citet{telles2014}, overlaid with the ALMA CO and millimeter-continuum contours from this work (same contour levels as in Fig.~\ref{fig-conts}). The maps provide spatially resolved diagnostics of the ionized gas excitation, temperature, density, and attenuation in the central region of \galaxy. The CO emission avoids the most highly ionized regions, highlighting the spatial separation between the molecular and ionized phases. The cross and cyan circle denote the location of the SSC.}
 \label{fig-ratios1}
\end{figure*}

This spatial segregation is further supported by the comparison between the CO and mm continuum emission with optical emission-line diagnostics in Fig.~\ref{fig-ratios1} from the Gemini/GMOS IFU data of \citet{telles2014}, which trace the physical conditions of the ionized gas.  
The ionized gas traced by H$\alpha$, 
peaks at the position of the SSC, where the nebular extinction (i.e., high $\ha$/$\hb$) and ionization parameter are higher (i.e., larger [O\,{\sc iii}] $\lambda\lambda4958,5007$/[O\,{\sc ii}] $\lambda\lambda3727,3729$), and the electron temperature and density are also higher (i.e., lower [O\,{\sc iii}] $\lambda\lambda4958,5007/\lambda4363$ and lower [S\,{\sc ii}] $\lambda6716$/[S\,{\sc ii}] $\lambda6731$). 
This nuclear region is also where most OB stars and Wolf--Rayet features are found \citep{james2009,telles2014}.  In contrast, the CO emission avoids these regions dominated by a strong radiation field and is preferentially located in more shielded environments, offset from the SSC.

The combination of these datasets robustly shows that the cold molecular gas traced by CO is not co-spatial with the most highly ionized gas. The observed offset is consistent with a scenario in which CO traces dense, shielded clumps surrounding the central starburst, while the molecular gas closer to the SSC is largely CO-dark due to photodissociation. This picture is supported by the strong UV radiation field and by evidence for significant dust and gas in the nuclear region (see Section~\ref{subsect-dusty_core}). In this context, the detected CO emission likely represents only a fraction of the total molecular gas, and caution is therefore required when using it to infer the full molecular gas budget in low-metallicity starbursts. Future observations of alternative molecular tracers, together with spatially resolved H$_2$ emission in the near- and mid-infrared, will be essential to test these scenarios in greater detail.

A similar decoupling between CO-bright and CO-dark molecular gas has been observed in other low-metallicity systems, where CO emission traces only the densest and best-shielded structures embedded within a more extended molecular reservoir \citep[e.g.,][]{Rubio2015,cormier2019}. However, compact starbursts hosting SSCs can exhibit different configurations. In systems such as NGC~5253 or Mrk~71-A, CO emission is co-spatial with the SSC and associated with very dense, embedded molecular gas \citep[e.g.,][]{Turner2015,Oey2017}.

\subsection{Gas distribution and kinematics: Implications for  feedback}\label{subsect-gasdiskin}

Optical spectroscopy shows that the ionized ISM in \galaxy\ is kinematically complex, with very broad components (${\rm FWHM}\sim 400$--$900~\kms$) in high-ionization lines and narrower components at larger scales \citep{thuan96,james2009,telles2014}. This indicates that the most extreme motions are confined to the highly ionized nuclear gas.

In contrast, the molecular gas traced by CO exhibits significantly different kinematic properties. The CO lines are relatively narrow (FWZI $\lesssim$ 80\,km\,s$^{-1}$), with velocity dispersions of a few to $\sim$\,25\,km\,s$^{-1}$, but clearly supersonic for cold molecular gas. In addition, the velocity field (Fig.~\ref{fig-moms}) shows a coherent gradient over scales of a few hundred parsecs and a systematic blueshift of the brightest emission with respect to the systemic velocity (Figs.~\ref{fig-dis_co10} and ~\ref{fig-dis_co21}), which may indicate a geometry in which a significant fraction of the molecular gas lies on the near side of the system. 
Despite these perturbations, there is no clear evidence for strong acceleration of the molecular gas. In particular, the absence of broad line wings and large velocity offsets contrasts with those in the ionized phase, which reaches velocities of several hundred km\,s$^{-1}$. 

At the same time, the CO profiles are not purely quiescent. In the brightest regions, asymmetric line profiles with low-level red tails extending up to $\sim$\,25\,km\,s$^{-1}$ are observed, which may indicate localized slow motions, such as expansion or weak flows. 
The measured velocity dispersions are consistent with turbulent gas in gravitationally bound or marginally bound molecular structures, as observed in giant molecular clouds \citep[e.g.,][]{Solomon1987}. In the case of \galaxy, the proximity to the SSC suggests that stellar feedback likely contributes to driving this turbulence. However, alternative contributions such as gravitational motions or local dynamical processes cannot be excluded.

The combination of moderate velocity dispersions, coherent gradients, and lack of strong high-velocity components supports a scenario in which the molecular gas is dynamically perturbed but resists efficient acceleration. This behavior is consistent with multiphase feedback models, where low-density gas is more easily accelerated, while dense molecular clumps are more resistant to entrainment \citep[e.g.,][]{Fluetsch2019, Veilleux2020}. 

Therefore, the CO kinematics are consistent with molecular gas that is only weakly dynamically coupled to feedback from the SSC. In this picture, the gas survives in dense structures that experience turbulence, compression, and mild bulk motions, but do not participate in outflows traced by the ionized phase. Overall, the molecular gas in \galaxy\ appears to occupy an intermediate regime, being neither dynamically quiescent nor strongly accelerated, but instead reflects the complex interaction between gravity, turbulence, and stellar feedback in a low-metallicity starburst environment.

\subsection{Nature and origin of the CO reservoirs}\label{subsect-nature_origin}

The multiwavelength view presented in Sect.~\ref{subsect-gasdiskin}, together with the CO excitation and flux recovery analysis (Sect.~\ref{subsect-sdish}), suggests that the molecular gas in \galaxy\ is not uniformly distributed, but instead organized into at least two distinct components. 

First, the nuclear region associated with the SSC is characterized by strong UV emission, significant nebular extinction, and bright millimeter continuum (Sect.~\ref{subsect-cont}). These properties indicate the presence of dust and gas in a compact environment. However, no significant CO emission is detected at this location. For a central aperture of 0\farcs8 in size ($\sim$90 pc), we measure a conservative 3$\sigma$ upper limit for CO(1-0) of 45 mJy kms$^{-1}$. This corresponds to an upper limit in molecular mass of 4.1$\times$ 10$^{6}$ M$_{\odot}$. This suggests that a substantial fraction of the molecular gas in the immediate vicinity of the SSC could be CO-dark, likely due to efficient photodissociation of CO molecules in the intense radiation field, while H$_2$ remains self-shielded \citep[e.g.,][]{wolfire2010,glover2011,madden2020}.

Second, the detected CO traces clumpy molecular structures located away from the SSC, preferentially in more obscured regions toward the northwest (Fig.\ref{fig-hst}). These CO-bright clumps are spatially associated with regions of higher optical extinction and lie outside the most strongly ionized zones. This configuration suggests that CO survives primarily in denser and more shielded environments, where the radiation field is sufficiently attenuated. 

This interpretation is consistent with observations of other low-metallicity dwarf galaxies, where CO emission is confined to small, dense cores embedded within a more extended molecular reservoir that is largely CO-dark \citep[e.g.,][]{Rubio2015,cormier2019}. 
In such systems, CO traces only a fraction of the total molecular gas, typically the highest column-density structures. 
In this context, the CO-bright clumps identified in our data likely represent the densest and most shielded peaks of a more extended molecular reservoir. Their relatively narrow linewidths and modest velocity dispersions (Sect.~\ref{sec:kinematics}) further suggest that these structures are gravitationally bound or marginally bound, and able to survive in the vicinity of the central starburst despite the strong feedback. 

An additional possibility is that part of the molecular gas has an external origin, e.g., through asymmetric inflow or minor accretion. Such processes could contribute to the buildup of dense, shielded gas structures and help explain the observed offset between the CO emission and the SSC. Similar scenarios have been proposed in other low-metallicity starbursts, such as NGC\,5253, where molecular gas is thought to be accreted or funneled toward the central regions \citep[e.g.,][]{Turner2015, Miura2018, Cormier2017}. However, the current data do not provide clear kinematic signatures of inflow (e.g., coherent non-circular streaming motions), and the observed velocity field can also be explained by locally perturbed gas dynamics. Therefore, while external gas supply remains a plausible contributing factor, it cannot be robustly established with the present observations.

Finally, an alternative, though more speculative, possibility is that part of the CO has re-formed in dense regions following an earlier phase of photodissociation \citep[e.g.,][]{wolfire2010,glover2011}. In low-metallicity environments, CO formation timescales can be longer than those of H$_2$, and CO may only appear once sufficiently high column densities are reached. However, the current data do not provide direct constraints to distinguish between selective survival and re-formation of CO, and both processes may contribute.

\subsection{Dense and dusty nuclear core}
\label{subsect-dusty_core}

The 115 and 230 GHz continuum emission peaks at the position of the SSC and decline sharply beyond the central $\sim$3\arcsec, indicating that the dust-dominated emission is strongly concentrated in the nuclear region. The brighter 230\,GHz continuum relative to 115\,GHz in the immediate vicinity of the SSC (Sect.~\ref{subsect-cont}) is consistent with warm dust heated by the starburst and/or modest optical-depth effects; a non-negligible free--free contribution is also expected within the central arcseconds. 

The nuclear dust concentration has a clear counterpart at shorter wavelengths. The Balmer decrement peaks at the SSC (Fig.~\ref{fig-ratios1}), implying substantial nebular extinction in the ionized gas, while the stellar continuum at the same position remains among the bluest in the field, dominated by the young massive population. This contrast suggests that the dust responsible for the nebular attenuation is mixed with (or lies in front of) the ionized gas on small scales, whereas the continuum colors largely reflect the intrinsic stellar population and are less sensitive to localized extinction in the central arcseconds. Such a configuration is expected in compact starbursts where the ionized gas is embedded in dusty material, but the observed optical continuum is dominated by unobscured sightlines to the SSC. Given the very young age of the SSC ($\le$5 Myr), it is plausible that the cluster remains partially embedded in its natal material, which would naturally account for the coexistence of strong nebular extinction and a comparatively blue stellar continuum.

On $\sim$100\,pc scales, the dust distribution becomes more asymmetric. The reddest region in the V--I color map lies to the northwest of the SSC (Fig.~\ref{fig-hst}) and coincides with the peak of the cold ISM traced by CO. This spatial offset between the dust-rich nucleus and the most strongly reddened continuum region highlights the role of geometry and selective shielding: the dust column toward the SSC is sufficient to attenuate the nebular gas, yet the stellar continuum remains dominated by the youngest population, while the largest continuum reddening traces denser, more opaque structures adjacent to the nucleus. Comparable decouplings between warm dust peaks, nebular extinction, and molecular/dust lanes are observed in other metal-poor systems hosting SSCs, where feedback and patchy obscuration produce strong anisotropies in the ISM \citep[e.g.,][]{Turner2015,Turner2017,Beck2018,Oey2017,Consiglio2016,Kepley2016}.

A plausible interpretation of the central region is that the SSC is embedded in a compact dusty environment, while the detected CO traces adjacent shielded structures where the cold ISM is preferentially concentrated.
Further progress will require observations that directly probe the dust-embedded phases of the ISM in the nucleus. Mid-infrared spectroscopy with \textit{JWST} can constrain the warm molecular component through H$_2$ lines, while far-infrared fine-structure lines and carbon tracers (e.g., [C\,\textsc{ii}]$158\, \mathrm{\mu m}$ and [C\,\textsc{i}]$370,609\, \mathrm{\mu m}$) can help characterize the energetics and column densities of the dusty gas that is only indirectly accessible at optical wavelengths.

\section{Conclusions}
\label{sect-concl}

We presented new ALMA observations of the CO(1–0) and CO(2–1) transitions in the low-metallicity blue compact dwarf galaxy \galaxy, complemented by optical and UV imaging from \textit{HST} and optical integral field spectroscopy from ground-based observations. These data provide a spatially resolved view of the molecular gas in an extreme, SSC-dominated, low-metallicity starburst. The main conclusions are as follows:

\begin{enumerate}

\item The CO emission is distributed in compact, clumpy structures within $\sim$800\,pc of the nucleus, and is spatially offset from the SSC and the most highly ionized regions. This indicates that the cold, CO-bright molecular gas phase is not co-spatial with the current sites of massive star formation, but instead resides in more shielded regions of the ISM.

\item The molecular gas exhibits relatively narrow CO line profiles  
with clearly supersonic velocity dispersions (a few to $\sim$25\,km\,s$^{-1}$ across both transitions) together with a coherent velocity gradient and a systematic blueshift ($\sim$\,20\kms) with respect to the systemic velocity. These properties indicate that the molecular gas is dynamically perturbed and participates in large-scale motions.

\item Despite these perturbations, there is no clear evidence for fast molecular outflows or strong acceleration of the CO-emitting gas. This contrasts with the ionized phase, which shows velocities of several hundred km\,s$^{-1}$. The molecular gas is consistent with being only weakly dynamically coupled to the feedback from the SSC, experiencing turbulence and mild bulk motions without being efficiently entrained.

\item The global CO(2--1)/CO(1--0) ratio is markedly low ($R_{21} \approx 0.29$), consistent with subthermal excitation and relatively low gas densities and/or beam dilution effects. Together with the partial recovery of the single-dish flux, this indicates that a significant fraction of the molecular gas is distributed in extended, low-surface-brightness structures that are not fully traced by the interferometric observations.

\item The millimeter continuum peaks at the SSC, while the CO emission is displaced toward regions of higher optical extinction. This suggests that the nuclear region hosts a compact, dusty, and likely CO-dark molecular component, while the detected CO traces denser clumps in shielded environments.

\item The molecular gas mass inferred from CO(1--0) is of order a few $10^{7}\,M_\odot$, although the absolute value remains uncertain at the $\sim$0.3--0.4 dex level due to the metallicity dependence of the CO-to-H$_2$ conversion factor.

\end{enumerate}

Our results support a multiphase picture of stellar feedback in which different components of the ISM respond in distinct ways. The highly ionized gas traces the most energetic and dynamically disturbed phase, while the molecular gas survives in dense, shielded structures that are only weakly affected by the feedback. Thus, CO traces only the densest and most shielded fraction of the molecular gas and likely misses a substantial CO-dark component. As a result, caution is required when using CO emission alone to infer the total molecular gas content and star formation efficiency in low-metallicity starbursts.

Our results highlight the importance of combining molecular, ionized, and dust tracers to obtain a complete view of the ISM in chemically primitive galaxies, and reinforce the need for alternative tracers of the molecular gas reservoir beyond CO in these environments. In this context, \galaxy\ illustrates how the observable CO emission may represent only the visible peaks of a much larger molecular reservoir shaped by feedback and low metallicity.

\begin{acknowledgements}
      This paper makes use of the following ALMA data: ADS/JAO.ALMA2021.1.01307.S ALMA is a partnership of ESO (representing its member states), NSF (USA) and NINS (Japan), together with NRC (Canada), NSTC and ASIAA (Taiwan), and KASI (Republic of Korea), in cooperation with the Republic of Chile. The Joint ALMA Observatory is operated by ESO, AUI/NRAO and NAOJ. 
      RS and RA acknowledge the support of ANID FONDECYT Postdoctoral Grant 3200909 and FONDECYT Regular Grant 1202007. RA acknowledges support of Grant PID2023-147386NB-I00 funded by MICIU/AEI/10.13039/501100011033 and by ERDF/EU, and the Severo Ochoa grant CEX2021-001131-S funded by MCIN/AEI/10.13039/501100011033. FJSR acknowledges financial support by PREP2023-001336 funded by MICIU/AEI/10.13039/501100011033 and ESF+. BLJ is thankful for support from the European Space Agency (ESA). MSO acknowledges the hospitality of the IAA/CSIC. JMV acknowledges financial support from the grants CEX2021-001131-S, funded by MICIU/AEI/10.13039/501100011033, and PID2022-136598NB-C32. JAFO acknowledges financial support by the Spanish Ministry of Science and Innovation (MCIN/AEI/10.13039/501100011033), by ``ERDF A way of making Europe'' and by ``European Union NextGenerationEU/PRTR'' through the grants PID2021-124918NB-C44 and CNS2023-145339; MCIN and the European Union -- NextGenerationEU through the Recovery and Resilience Facility project ICTS-MRR-2021-03-CEFCA. This work has rendered the images using the software tool CARTA. \citep{comrie2021}.
\end{acknowledgements}
%
   \bibliographystyle{aa} 
   \bibliography{bibmrk996} 
%



\begin{appendix}
\section{ALMA data calibration and imaging tests}
\label{app:alma_cal}

\subsection{Pipeline calibration and flagging}

The ALMA data for \galaxy\ were processed using the standard pipeline within \textsc{CASA}\footnote{http://casa.nrao.edu} versions~6.2.1.7 and~6.5.5.21 \citep{mcmullin2007}.
The delivered measurement sets included automatic flagging of shadowed antennas, bad baselines, and edge channels.
In addition, antenna DA64 was manually flagged in the CO(1–0) dataset owing to anomalous phase behavior during the December~2021 execution block.
No significant phase drifts or amplitude instabilities were found in the remaining antennas.
Bandpass, flux, and phase calibration followed the standard ALMA scripts, using J0006$-$0623 (bandpass/flux) and J0110$-$0741 and J0141$-$0928 (phase) as calibrators.
The typical amplitude accuracy after calibration is estimated to be better than 5\%.

\subsection{Imaging parameters and weighting tests}

Imaging was performed using the \texttt{tclean} task with a Briggs weighting scheme and different \texttt{robust} values between $-$1.0 and $+1.0$ to optimize the compromise between sensitivity and resolution.
Uniform and natural weighting were also tested for comparison.
A \texttt{robust}=0.5 combined with a Gaussian $uv$-taper (matched to $\sim$15\% of the beam) provided the best compromise, enhancing diffuse emission without significant beam elongation.
Table~\ref{tab:imaging_params} summarizes the adopted parameters for the final datacubes.

Primary-beam correction was applied to all cubes using \texttt{impbcor}.
The spectral resolution was kept at the native channel spacing to preserve the intrinsic linewidths of the narrow CO profiles.
We verified that moderate spectral binning (up to 5 channels) did not significantly alter integrated fluxes ($<$5\%).

\subsection{Continuum subtraction and moment masking}
\label{app:subsec-sub_mask}

Continuum subtraction was performed in the image domain using the \texttt{imcontsub} task, yielding more stable results than the $uv$-plane subtraction due to the weakness of the continuum relative to the CO lines.
Continuum maps were produced using multifrequency synthesis (\texttt{mfs}) and all line-free channels within the spectral windows.

Moment maps were created using the CASA task \texttt{immoments} after applying primary-beam correction and masking. To minimize noise, different masking strategies were applied to each transition according to their signal-to-noise properties. For CO(1–0), pixels below 1.6 mJy ($\sim$1$\sigma$) were excluded, and an additional mask was applied to the moment-0th map using 75 mJy beam$^{-1}$ km s$^{-1}$ ($\sim$3$\sigma$). For CO(2–1), pixels with intensities below 9.2 mJy ($\sim$4$\sigma$) were excluded prior to the calculation of the moments, and the higher SNR allowed no additional masking on the moment-0th map.
These thresholds were determined by inspecting line-free regions to ensure that noise peaks do not contribute to integrated flux.
Changing the mask by $\pm$1$\sigma$ affects the total integrated CO fluxes by less than 8\%, confirming the robustness of the measurements.

To derive the $R_{21}$ ratio, the CO(1–0) and CO(2–1) maps must first be placed on a common pixel grid. The maps have slightly different pixel sizes (0\farcs1 and 0\farcs14, respectively), which prevents direct pixel-by-pixel operations. The CO(2–1) moment-0 map was therefore regridded to the CO(1–0) pixel scale using the \textit{imregrid} task in \textsc{CASA}. This procedure does not improve the angular resolution, which remains set by the synthesized beam, but minimizes flux losses that become more significant when degrading the maps to coarser pixel scales.

The aligned maps were then intersected using a custom script to retain only the pixels with overlapping emission, masking all other regions. From these matched maps we measured the integrated flux of each transition, from which the CO luminosities and the corresponding $R_{21}$ value were derived (see Sect.~\ref{subsect-r21}).

\subsection{Flux consistency and recoverable scales}
\label{app:subsec-fluxMRS}

To assess possible flux losses, we integrated the visibilities over the $uv$-range corresponding to the IRAM~30\,m and APEX beams. According to the measured values in units of K \kms, we performed a unit conversion to Jy \kms, in order to facilitate comparisons and following the standard relation for transforming brightness temperature into integrated flux density:
\begin{equation}
    S\Delta\nu=\frac{2k_{B}\nu_{obs}^2}{c^2}I_{T_{B}}\Omega \times10^{26} ,
\end{equation}
where, k$_{B}$ is the Boltzmann constant (J/K); $c$ is the speed of light (ms$^{-1}$); $\nu_{obs}$ is the observed frequency (Hz); I$_{T_{B}}$ is the integrated flux in K \kms to be transformed, and $\Omega$ is the solid angle, estimated as $1.133\times \theta^2$, with $\theta$ being the beam size in radians \citep[see a similar conversion in][ Appendix A]{hunt2017}. The Table~\ref{table-co_compare}, summarizes these transformed values of total CO flux, along with those derived from our ALMA observations.

The recovered fluxes represent $\sim$42\% (CO(1–0)) and $\sim$50\% (CO(2–1)) of the single-dish values from \citet{hunt2015,hunt2017}.
This difference is consistent with the maximum recoverable scale (MRS) of $\sim$7$^{\prime\prime}$ in our interferometric configuration, implying that extended, diffuse emission is partly resolved out. 
The estimated flux recovery within the sampled spatial scales is consistent with unity, indicating that the measured emission within these regions is robust. 
After correcting for this effect, the global CO(2–1)/CO(1–0) ratio decreases from 0.34 to $\sim$0.29 (Sect.~\ref{sect-discuss}). While this correction aims to account for spatial effects, the resulting R$_{21}^{\rm corr}$ is significantly lower than the values typically reported in single-dish studies. In nearby galaxies, CO emission typically arises from optically thick molecular gas, with line ratios $R_{21}$ approaching unity in central regions and decreasing to $\sim$0.6–0.8 in outer disks of spiral galaxies \citep[e.g.,][]{leroy2009,saintonge2017,denbrok2021,egusa2022, keenan2024}. For dwarf galaxies, \citet{albrecht2004} measured CO(1–0) and CO(2–1) emission using the IRAM 30 m Telescope and the Swedish–ESO Submillimetre Telescope with matched $\sim$24\arcsec\ beams, allowing a direct estimate of $R_{21}$. Their velocity-integrated intensities yield ratios spanning $\sim$0.4–1.3 (see their Table 5). Similar values ($R_{21}\sim0.6$–1.3) have also been reported in starburst dwarf galaxies and giant molecular clouds in starburst environments \citep[e.g.,][]{taylor99}.

\subsection{Verification of positional alignment}

The astrometry of the ALMA images was verified by comparing the peak positions of the 230\,GHz continuum with the \textit{HST}/WFPC2 F569W images, resulting in offsets $<0\farcs1$.
The GMOS/IFU cube was reprojected onto the ALMA astrometric grid using common point-like sources in the V-band image as references.
Residual uncertainties in the alignment are estimated to be below 0\farcs15, negligible relative to the synthesized beams.

\begin{table}[t!]
\centering
\caption{Summary of imaging parameters and resulting beam characteristics.}
\label{tab:imaging_params}
\begin{tabular}{lcc}
\hline\hline
Parameter & CO(1–0) & CO(2–1) \\
\hline
CASA version & 6.5.5.21 & 6.5.5.21 \\
Weighting scheme & Briggs & Briggs \\
Robust parameter & 0.5 & 0.5 \\
UV-taper (arcsec) & 0.8 & 1.2 \\
Pixel scale (arcsec) & 0.12 & 0.08 \\
Synthesized beam & 0\farcs87$\times$0\farcs80 & 1\farcs21$\times$1\farcs04 \\
Beam PA (°) & 45.5 & 73.1 \\
RMS noise (mJy\,beam$^{-1}$) & 1.6 & 2.3 \\
Velocity channel (km\,s$^{-1}$) & 2.6 & 1.27 \\
\hline
\end{tabular}
\end{table}

\section{Dependence of molecular mass on the CO--\texorpdfstring{H$_2$}{H2} conversion factor}
\label{app:alphaCO}

In this section, we will test our molecular mass estimates against different metallicity-dependent prescriptions for $\alpha_{\rm CO}$. We compute the CO luminosity as in Section~3 (Eqs.~1 and 2). 

Unless stated otherwise, we adopt $12+\log({\rm O/H})=7.94$ ($\sim$0.2\,Z$_\odot$) for \galaxy\ and use the total CO(1–0) luminosity measured over the ALMA FOV,
$L'_{\rm CO(1–0)}=(6.5\pm1.0)\times10^{5}$ K\,km\,s$^{-1}$\,pc$^2$
(Section~3).

We test four widely used prescriptions:

\begin{enumerate}
\item \textbf{Amorín et al. (2016):}
\[
\log\alpha_{\rm CO}=0.7-1.5\left[(12+\log{\rm O/H})-8.7\right]
\;\Rightarrow\;
\alpha_{\rm CO}\approx 69.
\]

\item \textbf{Bolatto et al. (2013):}
\[
\alpha_{\rm CO}=\alpha_{\rm CO,MW}\left(\frac{Z}{Z_\odot}\right)^{-\gamma},
\quad \gamma\simeq1.6
\;\Rightarrow\;
\alpha_{\rm CO}\approx 57
\;\;{\rm at}\;\; Z\simeq0.2\,Z_\odot.
\]

\item \textbf{Schruba et al. (2012):}
a steeper metallicity trend at low $Z$:
\[
\alpha_{\rm CO}=\alpha_{\rm CO,MW}\left(\frac{Z}{Z_\odot}\right)^{-2.0}
\;\Rightarrow\;
\alpha_{\rm CO}\approx 109
\;\;{\rm at}\;\; Z\simeq0.2\,Z_\odot.
\]

\item \textbf{Hunt et al. (2015):}
a high-$\alpha_{\rm CO}$ scaling consistent with metal-poor dwarfs; we adopt
\[
\alpha_{\rm CO}=\alpha_{\rm CO,MW}\left(\frac{Z}{Z_\odot}\right)^{-1.9}
\;\Rightarrow\;
\alpha_{\rm CO}\approx 93
\;\;{\rm at}\;\; Z\simeq0.2\,Z_\odot.
\]
\end{enumerate}

These values are representative of each prescription at Mrk\,996's metallicity and illustrate the systematic spread ($\sim$0.3–0.4\,dex) inherent to low-$Z$ calibrations. Errors reflect the statistical uncertainty in $L'_{\rm CO}$. Systematic uncertainties from $\alpha_{\rm CO}$ dominate ($\sim$0.3–0.4\,dex).

For completeness, repeating the exercise in the overlapping region used to derive $R_{21}$ (Section~3; $L'_{\rm CO(1–0)}=(2.8\pm0.4)\times10^{5}$) gives:
\[
M_{\rm H_2}\,(\times10^{7}\,M_\odot) = [1.93,\,1.60,\,3.05\,2.60]
\]
for prescriptions 1 to 4, respectively. 
The choice of metallicity matters and varying $Z$ within $0.15$–$0.25\,Z_\odot$ yields:
\[
\alpha_{\rm CO}\;{\rm (Bolatto13)}\;\approx
40\text{–}91,\quad
\alpha_{\rm CO}\;{\rm (Schruba12)}\;\approx
70\text{–}174,
\]
which translates into a $\sim$2–4$\times$ range in $M_{\rm H_2}$ at fixed $L'_{\rm CO}$. This systematic spread should be considered when comparing to single-dish estimates or to galaxies with different metallicities.

\medskip
In summary, all prescriptions agree that \galaxy\ hosts a modest molecular reservoir at low excitation, but the absolute $M_{\rm H_2}$ is calibration-limited: $M_{\rm H_2}\sim(3.7$–$7.1)\times10^{7}\,M_\odot$ for the ALMA FOV, with \citet{amorin2016} giving an intermediate value consistent with $\sim$0.2\,Z$_\odot$ systems.

\begin{table}[!t]
\centering
\caption{Molecular gas masses from different $\alpha_{\rm CO}$ prescriptions (total ALMA FOV).}
\label{tab:alpha_tests_total}
\begin{tabular}{lcc}
\hline\hline
Prescription & $\alpha_{\rm CO}$ & $M_{\rm H_2}=\alpha_{\rm CO} L'_{\rm CO}$ ($10^{7}\,M_\odot$)\\
\hline
Amorín+2016 & 69  & $4.49 \pm 0.69$ \\
Bolatto+2013 ($\gamma{=}1.6$) & 57  & $3.71 \pm 0.57$ \\
Hunt+2015 ($\gamma{=}1.9$)    & 93  & $6.05 \pm 0.93$ \\
Schruba+2012 ($\gamma{=}2.0$) & 109 & $7.09 \pm 1.09$ \\
\hline
\end{tabular}
\end{table}

\end{appendix}

\end{document}